\documentclass[aps,pra,longbibliography,twocolumn,showpacs,eqsecnum,amsmath,amssymb,floatfix,superscriptaddress]{revtex4-1}
\usepackage{graphicx} 
\usepackage{dcolumn} 
\usepackage{bm} 
\usepackage{hyperref} 
\usepackage{slashed}
\usepackage{subfigure}

\def\beq{\begin{equation}}  
\def\eeq{\end{equation}}

\def\ben{\begin{align}}
\def\een{\end{align}}

\begin{document}

\title{Theory of Nematic Fractional Quantum Hall State}

\author{Yizhi You}
\affiliation{Department of Physics and Institute for Condensed Matter Theory,
University of Illinois at Urbana-Champaign, 1110 West Green Street, Urbana, Illinois 61801-3080, USA}
\author{Gil Young Cho}
\affiliation{Department of Physics and Institute for Condensed Matter Theory,
University of Illinois at Urbana-Champaign, 1110 West Green Street, Urbana, Illinois 61801-3080, USA}
\author{Eduardo Fradkin}
\affiliation{Department of Physics and Institute for Condensed Matter Theory,
University of Illinois at Urbana-Champaign, 1110 West Green Street, Urbana, Illinois 61801-3080, USA}
\affiliation{Kavli Institute for Theoretical Physics, University of California Santa Barbara, CA 93106-4030, USA}

\date{\today}

\begin{abstract}
We derive an effective field theory for the isotropic-nematic quantum phase transition of fractional quantum Hall (FQH) states. We demonstrate that for a system with an isotropic background  the low-energy effective theory of the nematic order parameter has $z=2$ dynamical scaling exponent, due to a Berry phase term of the order parameter, which is related to the non-dissipative Hall viscosity. Employing the composite fermion theory with a quadrupolar interaction between electrons, we show that  a sufficiently attractive quadrupolar interaction triggers a phase transition from the isotropic FQH fluid into a nematic fractional quantum Hall phase. By investigating the spectrum of collective excitations, we demonstrate that the mass gap of Girvin-MacDonald-Platzman (GMP) mode collapses at the isotropic-nematic quantum phase transition. On the other hand, Laughlin quasiparticles and the Kohn collective mode remain gapped at this quantum phase transition, and   Kohn's theorem is satisfied. The leading couplings between the nematic order parameter and the gauge fields include a term  of the same form as the Wen-Zee term. A disclination  of the nematic order parameter carries an unquantized electric charge.  We also discuss the relation between nematic degrees of freedom and the geometrical response of the fractional quantum Hall fluid. 
\end{abstract}

\maketitle

\section{Motivation and Introduction}

Strongly correlated electronic systems have a strong tendency to have liquid-crystal-like ground states (e.g. crystals, smectics or stripes, and nematics)  which break spontaneously translation and rotational invariance to varying degrees  \cite{kivelson-1998}. Typically, these states  arise as the result of the competition between repulsive Coulomb interactions and effective attractive interactions that arise from the disruption of strongly correlated states in systems with microscopic repulsive interactions.
In two-dimensional electron gases (2DEGs) in large magnetic fields these effects  are even stronger since the kinetic energy of the electron is completely quenched in  an uniform perpendicular magnetic field and hence  interaction effects are dominant. For these reasons, in addition to incompressible quantum Hall states (FQH), integer or fractional, electronic liquid crystal phases are generally expected to occur in these systems  \cite{Fradkin-1999}.

Theoretically, several Hartree-Fock studies  \cite{koulakov-1996,koulakov-1996b,macdonald-2000,Fogler2002} (and effective field theories  \cite{barci-2002}) have predicted stripe phases, as well as ``bubble'' and other crystalline states  \cite{Lee2002}, in addition to the expected Wigner crystals  \cite{Kivelson1986PRL,Kivelson1987PRB,Tesanovic1989,Prange-1990,Murthy2000}.
These phases  are expected to become exact ground states for very weak magnetic fields  \cite{moessner-1996}, and in effective field theories \cite{barci-2002} Similarly, (compressible) nematic phases have been found in variational wave-function calculations \cite{doan-2007,wexler-2001,wexler-2004} and also in phenomenological hydrodynamic theories  \cite{Radzihovsky2002}. Exact diagonalization studies of small systems have found evidence of short-range stripe order in a Landau level  \cite{rezayi-2000}.
 For a recent review on electronic nematic phases see Ref.\cite{Fradkin-2010}. 

Experiments in the second Landau level, $N=2$ (and in the first Landau level, $N=1$, in tilted fields) have established the existence of {\em compressible} states of the 2DEG with an extremely large transport spatial anisotropy with a marked temperature dependence  \cite{lilly-1999,du-1999,cooper-2002}, a {\em nematic Fermi fluid}  \cite{Fradkin-1999,Fradkin-2000}. In these experiments the anisotropy probed by a small in-plane component of the magnetic field which breaks rotational invariance explicitly. However the strong temperature dependence of the anisotropy implies that that the in-plane field reveals a strong tendency to break rotational invariance spontaneously. Thus, the measured anisotropy of the transport can be regarded as the response to the in-plane field exactly in the same way as the magnetization is the response to a Zeeman field in a magnet. In this sense the anisotropy vs in plane field curves can be regarded as the equation of state of the 2DEG (or, rather, the nematic susceptibility). On the other hand, given the absence of pinning effects observed in this regime,  the linearity of their $I-V$ curves at low voltages, and the scaling behavior exhibited by the data, one can readily conclude that  these states are regarded as  (compressible) electron nematic states  \cite{Fradkin-2000} rather than stripes (or unidirectional charge density waves (CDW)), or ``bubble'' phases (i.e. multi-directional CDW states), expected from Hartree-Fock calculations
  \cite{koulakov-1996,koulakov-1996b,macdonald-2000,Fogler2002}. To this date, the compressible nematic state in the $N=2$ Landau level near filling fraction $\nu=9/2$ is the best documented case of an  nematic phase in any electronic system  \cite{Fradkin-2010}.

More recent magneto-transport experiments in the first, $N=1$, Landau level, have shown that  {\em incompressible} fractional quantum Hall state with filling fraction $\nu=7/3$ have a pronounced temperature-dependent anisotropy in their longitudinal transport. As in all experiments of this type (see, e.g., the review of Ref. \cite{Fradkin-2010}) the anisotropy is seen in the presence of a weak symmetry-breaking field (here, the in-plane component of the magnetic field) which reveals a pronounced (but smooth) rise of the transport anisotropy as the temperature is lowered below some characteristic value. Since the symmetry is broken explicitly, these experiments provide 
evidence for a large temperature-dependent {\em nematic susceptibility} in these fluid states  \cite{Xia2011}. These experiments strongly suggest that, at least in the $N=1$ Landau level, the FQH phases the 2DEG may be close to a phase transition to an {\em incompressible} nematic state  {\em inside} the topological  fluid phase, i.e. a nematic FQH state.
The notion of a  nematic FQH state was actually suggested early on by Balents \cite{Balents1996}. However, this  concept did not attract significant attention until  the recent experiments of Xia and coworkers which suggested the existence of strong nematic correlations became available \cite{Xia2011}. 

The experiments of Xia and coworkers motivated Mulligan and coworkers \cite{Mulligan2010,Mulligan2011} to formulate a theory of a quantum phase transition inside the $\nu=7/3$ FQH phase, from an isotropic fluid to a nematic FQH state interpreted as a quantum Lifshitz transition. The theory of Mulligan {\it et al.} uses as a starting point the effective field theory of a Laughlin isotropic FQH fluid with filling fraction $\nu=1/m$ (with $m$ an odd integer) whose effective Lagrangian is that of a (hydrodynamic) gauge field, $a_\mu$, with a Maxwell and a Chern-Simons term \cite{Wen-1995,fradkin-1991}. In this picture, the FQH quantum Lifshitz transition occurs when the coefficient of the electric field term of the Maxwell-like term of the effective action of the hydrodynamic gauge field vanishes, and can be regarded as a Chern-Simons version of the quantum Lifshitz model \cite{Ardonne-2004}.

While this theory successfully predicts many aspects of the experiment (in particular the anisotropy) it has several difficulties, the most serious of which is that in a Galilean invariant system the coefficient of the term for hydrodynamic electric field is fixed by Kohn's theorem \cite{kohn-1961}. Although this restriction can be violated by a relatively small amount by Landau level mixing effects \cite{Mulligan2011}, it is unlikely to become large enough to trigger a Lifshitz transition to a nematic state. Another puzzling aspect is that the Chern-Simons Lifshitz theory of Mulligan {\it et al.} also applies to the integer Hall states. However, barring large enough Landau level mixing effects, it is hard to see how a system in the integer quantum Hall  regime may break spontaneously rotational invariance. The experiment of Xia {\it et al} have also prompted several studies of integer and fractional quantum Hall states in systems in which the anisotropy is built-in explicitly in the geometry of the two dimensional surface in which the electrons reside \cite{Yang-2013}, including wave functions for states with  fixed anisotropy \cite{Qiu-2012}.

Maciejko and coworkers \cite{Maciejko2013} recently proposed an effective field theory of the spontaneous breaking of rotational invariance in a nematic state in the FQH regime with the form of a non-linear sigma model on the non-compact target space $SO(2,1)^+$ manifold of the rotational degrees of freedom and the amplitude of the local nematic order parameter. They proposed that the nematic transition is triggered by a softening of the intra-Landau level Girvin-MacDonald-Platzman (GMP) collective mode of the FQH fluid \cite{gmp}. A key result from this work is the observation that, due to the breaking of time-reversal invariance in the FQH fluid,  the dynamics of the nematic fluctuations is governed by a Berry phase term, whose coefficient they conjectured to be essentially the same as the (non-dissipative) Hall viscosity of the FQH fluid \cite{Avron1995,Read2009,Read2011}. Maciejko and coworkers also further an interpretation of nematic fluctuations as a fluctuating geometry (making contact with ideas put forward by Haldane on the existence of geometric degrees of freedom in the FQH liquid \cite{Haldane2009,Haldane2011,Yang-2013}). Similar ideas were discussed by two of us in the context of a nematic transition in a spontaneous anomalous quantum Hall state \cite{you}, and earlier on by one of us in a  theory of thermal melting of the pair-density-wave superconducting state \cite{barci-2011}. The conjectured connection between the nematic fluctuations in the FQH fluid and the Hall viscosity strongly suggest a relation with theories of the geometric response of these topological fluids \cite{son-2013,Abanov2014,Gromov2014,Cho-2014,bradlyn-2014}, which we will further elaborate below.

In this paper we address several open aspects of this problem that have remained unexplained. One of the issues is the origin of the  nematic quantum phase transition which Maciejko {\it et al.} argued could be due to a softening of the GMP collective mode. Here we will show that the GMP mode can become gapless at wave vector ${\bm q}=0$ if the {\em effective interactions} among the {\em electrons} are sufficiently attractive in the {\em quadrupolar} channel. It is known that in a  Fermi liquid, a sufficiently attractive effective interaction in the quadrupolar channel (i.e. a sufficiently negative charge-channel Landau parameter $F_2$) can trigger a nematic instability through a Pomeranchuk instability which results in a spontaneous quadrupolar distortion of the Fermi surface \cite{oganesyan-2001}. Here we will postulate that at long wavelengths, in addition to the long-range Coulomb interaction, there is an attractive short-range quadrupolar interaction. Such an effective interaction can arise due to the softening of the short-distance Coulomb interaction in Landau levels $N\geq 1$. In fact, an early numerical study by Scarola and coworkers \cite{Scarola2000} of the effective interactions of {\em composite fermions} \cite{Jain1989} showed that in Landau levels with $N\geq 1$ there is a strong tendency for the FQH liquid to become unstable (and was interpreted as an exciton instability.) From the point of view of symmetry breaking, a ${\bm q}=0$ (`exciton') quadrupolar condensate is equivalent to an instability to nematic state since they break the same spatial symmetries. 
The other focus of this work is to clarify the relation between the nematic fluctuations (and possible order) in the FQH fluid to the response of this fluid to changes on the actual background geometry of the surface on which the 2DEG resides. This is an important question since quantities such as the Hall viscosity measures the response to shear deformations of the geometry and this is not quite the same as the  the nematic response, although, as we will see below, they are related. 

In order to study the quantum nematic phase transition in a FQH fluid we first generalize the fermion Chern-Simons theory of the  FQH states \cite{Lopez-1991} to include the effects of the attractive quadrupolar interaction, and show that  indeed there can be a quantum phase transition inside all Jain states of the FQH provided the quadrupolar interaction is sufficiently attractive. In our treatment we  also include the coupling to the background geometry of the 2D surface on which the 2DEG resides. We then use our recent results presented in Ref. \cite{Cho-2014} to show that the quadrupolar interaction couples to both the so-called statistical gauge field (of the fermion Chern-Simons theory) and to the spin connection of the geometry. Our first main result is the derivation of the effective action for the nematic degrees of freedom which, as expected, has the form proposed by Maciejko {\it et al.}. The fluctuations of the nematic order parameter are  strongly coupled to the GMP mode of the FQH fluid (which has quadrupolar character), and the nematic quantum phase transition is triggered when the ${\bm q}=0$ component of  this mode becomes gapless.
Furthermore, the dynamics of the nematic degrees of freedom is controlled by a Berry phase term and, hence, has dynamics critical exponent $z=2$.  However its coefficient is not the Hall viscosity of the FQH fluid (as conjectured in Ref. \cite{Maciejko2013}) but is given, instead, by the Hall viscosity of the effective integer Hall effect of the composite fermions. Nevertheless, the Hall viscosity of the system (both in the isotropic and in the nematic phase), defined as the response to the shear deformation of the underlying geometry, is the same as the Hall viscosity of the FQH fluid obtained in Ref.\cite{Read2009} (and recently rederived by us \cite{Cho-2014}). These results are reminiscent of the previous study by two of us~ \cite{you} where we have studied the effective theory of the phase transition between an isotropic Chern insulator and a nematic Chern insulator.
We  also demonstrate that in this theory the nematic transition is reached while the Kohn mode remains unaffected  in both phases and at the phase transition. In addition we also show that the components of the nematic order parameter can be used to define an effective spin connection (which is effectively the same as the ``nematic gauge field'' phenomenologically introduced in Ref. \cite{Maciejko2013})
and that it couples to an external electromagnetic probe field through a term with the form of  the Wen-Zee term  \cite{Wen_Zee}, a result also anticipated by Maciejko {\it et al.}  We also derive the effective action for the spin connection of the background geometry and show that has the same form (with the same universal coefficients) in both phases. Finally we use our effective field theory to investigate the properties of a disclination of the nematic order parameter in the nematic phase, and show that it carries a fractional (but non-universal) electric charge and that the Hall viscosity is modified by the disclination, which agrees with the a symmetry-based argument of Ref.\cite{Maciejko2013}.

This paper is organized as follows. The theory of spontaneous rotational symmetry breaking is developed in Section \ref{SRSB}. After summarizing the fermion Chern-Simons gauge theory in Subsection \ref{FCS}, in Subsection \ref{quadrupolar} we introduce the quadrupolar interaction and its coupling with the statistical gauge field and with the spin connection of the background metric. In Section \ref{LGNematic} we derive the effective Landau-Ginzburg theory from the fermion Chern-Simons theory and in Section \ref{Nematic-QPT} we show that there is a  quantum phase transition to a nematic phase for sufficiently strong attractive quadrupolar coupling. In Section \ref{sec:disclinations} we discuss the behavior of the Goldstone mode of the broken orientational symmetry in the nematic phase and the nature of the disclinations. The coupling to the background geometry is developed in Section \ref{geometric-FQH}.
 Our conclusions are presented in Section \ref{conclusions}. In several appendices we present details of the calculation of the effective field theory. In Appendix \ref{app:calculation} we present the calculation of the nematic correlators and in Appendix \ref{app:mixed-correlators} the calculation of  mixed correlators of nematic  and gauge fields. A proof of gauge invariance is given in Appendix \ref{app:gauge-invariance}, and  the nematic collective excitations  are derived in Appendix \ref{app:collexcitation}.

\section{Spontaneous breaking of rotational symmetry in FQH states}
\label{SRSB}

\subsection{Composite Fermion Theory of FQH states}
\label{FCS}

Here we begin with a short review of the composite fermion theory of a FQH state \cite{Lopez-1991, Jain1989}, specializing in the simpler case of the Laughlin state at filling $\nu = \frac{1}{3}$, which can be easily generalized to the other states in Jain sequence $\nu = p/(2sp + 1)$, where $ s, p \in \mathbb{Z}$. Let us consider a theory of electron field $\Psi$ in two space dimensions in an uniform magnetic field. The action for this system is
\begin{align}
\label{original}
\mathcal{S} =& \int d^{2}x dt \left[ \Psi^{\dagger}(x) D_{0} \Psi(x) - \frac{1}{2m_e} (\bm{D}\Psi (x))^{\dagger}\cdot (\bm{D}\Psi (x))\right] \nonumber \\
 -&\frac{1}{2}\int d^{2}x'd^{2}xdt \; V(|\bm{x-}\bm{x}'|) \Psi^{\dagger}(x)  \Psi(x)  \Psi^{\dagger}(x') \Psi(x') 
\end{align}
 in which $D_{\mu} = \partial_{\mu} + i  A_{\mu}$ is the covariant derivative of the electron,  
 $m_e$  is the mass of the electron, and we have set  the Planck constant $\hbar$, the speed of light $c$, and the electric charge $e$ to unity. 
 The four-fermion term encodes the two-body interaction between the electrons. The electromagnetic gauge field $A_{\mu}$ can be written as $A_{\mu} = {\bar A}_{\mu} + \delta A_{\mu}$ where ${\bar A}_{\mu}$ is for the uniform magnetic field ${\bar B} = \epsilon^{ij} \partial_{i} {\bar A}_{j}$ perpendicular to the plane and $\delta A_{\mu}$ is the probe field to measure the response of the FQH state. 
 
 The average electron density ${\bar \rho}$ and the uniform external magnetic field ${\bar B}$ are related to each other through the filling fraction $\nu$
\beq
{\bar \rho} = \frac{\nu}{2\pi} {\bar B} = \frac{1}{6\pi} {\bar B}
\eeq
where we have set $\nu=1/3$ for the leaden Laughlin state. For a general Jain state the filling fraction is $\nu=p/(2sp+1)$, where $s$ and $p$ are two integers.
The Laughlin  FQH state with $\nu=1/3$ can be pictorially understood as the liquid state of the electrons in which, on average, each electron is bound with the two flux quanta. For a general Jain state, each electron is bound to $2s$ flux quanta and becomes a composite fermion \cite{Jain1989}.

This is a problem of strongly coupled electrons and  cannot tackled  directly using weak coupling methods.
To make progress, we follow Ref.\cite{Lopez-1991} and consider the equivalent system obtained by coupling the system of interacting electrons to  the (dynamical) Chern-Simons term of the statistical gauge field $a_{\mu}$, using minimal coupling. (For a detailed discussion see Ref.\cite{fradkin-1991}.) The action action of the equivalent problem is
\begin{align} 
\mathcal{S} =& \int d^{2}x dt \left[ \Psi^{\dagger}(x) D_{0} \Psi(x) - \frac{1}{2m_e} (\bm{D}\Psi (x))^{\dagger}\cdot (\bm{D}\Psi (x))\right]  \nonumber \\
-&\frac{1}{2}\int d^{2}x'd^{2}xdt \left[ V(|{\bm x}-{\bm x}'|) \Psi^{\dagger}(x)  \Psi(x)  \Psi^{\dagger}(x') \Psi(x')\right] \nonumber \\
+&\frac{1}{8\pi} \int d^{2}x dt \; \epsilon^{\mu \nu \lambda} a_{\mu}\partial_{\nu}a_{\lambda} 
\label{eq:FCS}
\end{align}
where $D_{\mu} = \partial_{\mu} + i A_{\mu} + i a_{\mu}$ is a new covariant derivative which includes the minimal coupling to both the electromagnetic field $A_\mu$ and to the statistical field $a_\mu$. This is the exact mapping of the original problem defined by the action of Eq. \eqref{original}. The Chern-Simons term binds the two flux quanta to the electron and turns the electron into the composite fermion \cite{Jain1989,jain-1992}.

We next consider uniform states which can be described using the average field approximation  in which we smear out the two flux quanta bound to the electron over the two-dimensional plane. This translates as choosing the average part of ${\bar a}_{\mu}$ to partially cancel the external magnetic field  ${\bar A}_{\mu}$. For the $\nu=1/3$ Laughlin state the effective field is
\beq
{\bar A}_{\mu} + {\bar a}_{\mu} = \frac{1}{3} {\bar A}_{\mu}
\eeq
Thus the composite fermion $\Psi$ is subject to the magnetic field which is $\frac{1}{3}$ of the magnetic field experienced by the electron. The composite fermion is in the integer quantum Hall effect at the filling $\nu =1$ and in effect is weakly coupled. We can write out the Lagrangian of the composite fermion. 
\begin{align}
\mathcal{S}  = &\int d^{2}x dt \left[ \Psi^{\dagger}(x) D_{0} \Psi(x) - \frac{1}{2m_e} (\bm{D}\Psi (x))^{\dagger}\cdot (\bm{D}\Psi (x))\right] \nonumber\\
 -&\frac{1}{2}\int d^{2}x'd^{2}x \left[ V(|{\bm x}-{\bm x}'|) \Psi^{\dagger}(x)  \Psi(x)  \Psi^{\dagger}(x') \Psi(x') \right] \nonumber\\ 
+&\frac{1}{8\pi}  \int d^{2}x dt \; \epsilon^{\mu\nu\lambda}\delta a_{\mu}\partial_{\nu} \delta a_{\lambda} 
\label{sec1sub1:CFT}
\end{align}  
Here  and below we denote by  $D_\mu$
\begin{equation}
D_{\mu} = \partial_{\mu} +i \frac{1}{3} {\bar A}_{\mu} + i\delta a_{\mu} +i \delta A_{\mu}
\label{eq:CD-CF}
\end{equation}
the covariant derivative of the composite fermion (again, for the $\nu=1/3$ Laughlin state). The fields $\delta a_{\mu}$ and $\delta A_{\mu}$ are the fluctuation of the gauge fields about their average values. Furthermore, the density fluctuation of the electron $\delta \rho = \Psi^{\dagger}\Psi - {\bar \rho}$ is bound with the flux of $\delta a_{\mu}$
\beq
\delta \rho (x) = \frac{1}{4\pi} \delta b(x) = \frac{1}{4\pi} \varepsilon^{ij} \partial_{i} \delta a_{j}
\eeq
This makes the density-density interaction between the electrons to be quadratic in the gauge field $\delta a_{\mu}$. As the action is quadratic in the composite fermion field $\Psi$, we can integrate out the fermion and the fluctuating part $\delta a_{\mu}$ of the statistical gauge field to obtain the effective theory for the gauge field $\delta A_{\mu}$. From the effective theory of $\delta A_{\mu}$, one can calculate the electromagnetic Hall response and find the collective excitations of the FQH state \cite{Lopez-1993} which, with some caveats,  agree qualitatively long wavelengths  with the experiments and numerical calculations.

\subsection{Quadrupolar Interaction}
\label{quadrupolar}

The composite fermion theory we just summarized is so far is rotationally invariant and cannot describe a nematic FQH state. Thus we should look for a new ingredient to the composite fermion theory to describe the nematic state and the transition toward the nematic state from the isotropic state. Since the density-density interaction of electrons and Chern-Simons term (at the level of the bare action of the composite fermion theory) cannot induce the spontaneous breaking of the rotational symmetry, we should look for an interaction which can favor the anisotropic state rather than the isotropic state. To this effect we follow the approach of the nematic Fermi fluid of Ref. \cite{oganesyan-2001} and add a quadrupolar interaction term $\mathcal{S}_q$ to the action of the form
\begin{equation}
\mathcal{S}_{q}=- \frac{1}{2}\int dt \int d^{2}x d^{2}x' F_2(|\bm{x}-\bm{x}'|) \textrm{Tr}[Q(x)Q(x')]
\label{eq:Sq}
\end{equation}
where  $F_2(|{\bm x}-{\bm x}'|)$ is the Landau interaction in the quadrupolar channel whose spatial Fourier transform is
\begin{equation}
F_2(\bm q)=\frac{F_2}{1+\kappa {\bm q}^2}
\label{eq:F2}
\end{equation}
and $\kappa>0$ parametrizes the interaction range.  
The coupling constant $F_2$ (i.e. the Landau parameter)  has units of energy $\times$ (length)$^6$. Here
we introduced the $2 \times 2$ traceless symmetric tensor  $Q(x)$
\begin{equation}
Q(x)=\Psi^{\dagger}(x)
\begin{pmatrix} 
 D^2_x-D^2_y  & D_x D_y+D_yD_x\\ 
  D_x D_y+D_yD_x &  D^2_y-D^2_x
\end{pmatrix}\Psi(x),
\label{sec1sub2:quadrupolar}
\end{equation}
 Here $D_x$ and $D_y$ are the spatial covariant derivatives defined in Eq.\eqref{eq:CD-CF}. 
 
The full action (including the quadrupolar interaction  $\mathcal{S}_q $) is manifestly rotationally invariant. In the case of a Fermi liquid, for large enough attractive quadrupolar interactions,  $F_2<0$, there is a Pomeranchuk instability which results in the spontaneous breaking of rotational invariance and the development of a {\em nematic} phase \cite{oganesyan-2001}. Here too, if $F_2<0$ and large enough in magnitude,  the quadrupolar coupling can induce a transition to an anisotropic phase by developing the finite expectation value of $Q(x)$. When $\langle  Q \rangle \neq 0$, the continuous rotational symmetry $O(2)$ of the two-dimensional space is broken down to $C_{2}$ generated by the discrete $\pi$ rotation of the plane. However, in the case of a Fermi fluid at zero external magnetic field the nematic phase leads to the spontaneous distortion of the Fermi surface and the development of an anisotropic effective mass for the quasiparticles in the anisotropic state. Furthermore, in the absence of a coupling to the underlying lattice the resulting nematic phase is a non-Fermi liquid. In the case at hand, although there is no Fermi surface to begin with, at the level of mean field theory, nematicity is also manifest as an effective anisotropy of the effective mass of the composite fermions.

Next, we include the quadrupolar interaction in the  the (fermionic) Chern-Simons theory of the FQH states  \cite{Lopez-1991} of Eq.\eqref{eq:FCS}
\begin{align}
\mathcal{S} =
 \int & d^{2}x dt \Big[ \Psi^{\dagger}(x) D_0 \Psi(x)-\frac{1}{2m_e}\left( {\bm D}\Psi (x) \right)^{\dagger} \cdot \left( {\bm D}\Psi(x) \right)\Big] 
\nonumber\\
-&\frac{1}{32\pi^{2}}\int d^{2}x'd^{2}x dt \;  V(|{\bm x}-{\bm x}'|) \delta b(x) \delta b(x')  \nonumber\\
+&\frac{1}{8\pi}  \int d^{2}x dt \; \epsilon^{\mu\nu\lambda} \delta a_{\mu}\partial_{\nu} \delta a_{\lambda} \nonumber\\
-&\frac{1}{2}\int dt \int d^{2}x d^{2}x' F_2(|\bm{x}-\bm{x}'|) \textrm{Tr}[Q(x)Q(x')]
\end{align}
Here we used the Chern-Simons constraint (i.e. the ``Gauss law'')  to represent the fluctuating density $\delta \rho$ of the composite fermion in terms of the fluctuating statistical field $\delta b$ which results in density-density interaction quadratic in the statistical gauge field. 

However, the quadrupolar interaction cannot be written as a  quadratic form in the statistical gauge field $\delta a_{\mu}$. Instead,  we perform a Hubbard-Stratonovich decoupling transformation to rewrite the quadrupolar interaction term $\mathcal{S}_q$ in terms of two fields $M_1$ and $M_2$ (which can be regarded as the two real components of a $2 \times 2$ real symmetric matrix field). After decoupling the action $\mathcal{S}_q$ of Eq.\eqref{eq:Sq} takes the form
\begin{align}
&\mathcal{S}_{q}= \int d^{2}x dt  \Big[  \frac{1}{4F_2m^2_e} {\bm M}^{2} - \frac{ \kappa}{4F_2m^2_e} \sum_{i=1,2} |{\bm \nabla} M_i|^2 \nonumber\\ 
&\frac{M_{1}}{m_e} \Psi^{\dagger}(D_x^2  - D_y^2) \Psi +  \frac{M_{2}}{m_e} \Psi^{\dagger} (D_x D_y + D_y D_x)\Psi \Big], 
\label{sec1sub2:HS}
\end{align}
Here we introduced suitable factors of the electron mass $m_e$ to make the Hubbard-Stratonovich fields $M_1$ and $M_2$ dimensionless. 
$F_2$ is the coupling constant of the quadrupolar interaction of Eq.\eqref{sec1sub2:quadrupolar}. 

It is apparent that in Eq.\eqref{sec1sub2:HS} $M_1$ and $M_2$ play the role of the order parameters for the nematic phase.  These fields  couple to the  the stress tensor tensor of the composite fermions and thus play a role analogous to a  background metric. In this sense, we can regard the nematic fluctuation as providing a ``dynamical metric'' which  modifies the local geometry of the composite fermions \cite{barci-2011,you}.

Thus we end up with the following action for the composite fermions coupled to the Chern-Simons gauge field, with a density-density interaction and a quadrupolar interaction,
\begin{align}
\mathcal{S} = \int & d^{2}x dt \left[ \Psi^{\dagger}(x) D_0 \Psi(x)-\frac{1}{2m_e}\left( {\bm D}\Psi (x) \right)^{\dagger} \cdot \left( {\bm D}\Psi(x) \right)\right] \nonumber \\
 -&\frac{1}{32 \pi^2}\int d^{2}x'd^{2}x dt \; V(|{\bm x}-{\bm x}'|) \delta b(x) \delta b(x')  \nonumber\\
+& \frac{1}{8\pi}\int d^{2}x dt \; \epsilon^{\mu\nu\lambda} \delta a_{\mu}\partial_{\nu} \delta a_{\lambda} \nonumber\\ 
+& \int d^{2}x dt  \Big[  \frac{1}{4F_2m^2_e} {\bm M}^{2} + \frac{ \kappa}{4F_2m^2_e} \sum_{i=1,2} |{\bm \nabla} M_i|^2 \nonumber\\ 
+& \frac{M_{1}}{m_e} \Psi^{\dagger}( D_x^2  -  D_y^2) \Psi + \frac{M_{2}}{m_e} \Psi^{\dagger} (D_x D_y + D_y D_x)\Psi \Big]
\label{sec1sub2:effective}
\end{align}
where, again,  the covariant derivatives are given in Eq.\eqref{eq:CD-CF}.
So far we have not made any approximations. In  the next section we will discuss the uniform states that result by treating this theory in the average field approximation and by considering the effects of fluctuations at the one loop (``RPA'') level.

It turns out  that from the theory we have defined the resulting quantum phase transition is strongly  first order and to a state with maximal nematicity (and without Landau quantization!). To avoid this pathological limit, and to make the nematic phase stable (and accessible by a continuous quantum phase transition), we will introduce an extra term in the kinetic energy part of the action of the form
\begin{equation}
S_6=-\alpha \int d^2x dt \Psi ^\dagger \left(\frac{-{\bm D}^2}{2m_e}-\frac{\bar \rho \pi}{m_e}\right)^3  \Psi
\label{eq:S4}
\end{equation}
where, once again, ${\bm D}$ stands for the space components of the (full) covariant derivative and ${\bm D}^2$ is the covariant Laplacian.
A term of a similar type was introduced by Oganesyan {\it et al.} \cite{oganesyan-2001} in their theory of the nematic Fermi fluid formed by a Pomeranchuk instability. Here too this (technically irrelevant) term will insure that the nematic state is stable, provided the coupling constant $\alpha$ is large enough (as we will see below). For other ranges of $\alpha$ the quantum phase transition becomes first order, as it happens in  theories of the electronic nematic transition in lattice systems \cite{khavkine-2004}. Although the addition of this term complicates the calculation somewhat, it does not change the physics in any essential way. We should note that this term commutes with the gauge-invariant kinetic energy and, consequently, it has the same eigenstates. Thus, this term changes only the eigenvalues but it does not induce Landau level mixing.

\subsection{Symmetries}
\label{sec:symmetries}

The action of Eq.\eqref{sec1sub2:effective} has two important symmetries. One is local gauge invariance, under which the Fermi field $\Psi(x)$ and the gauge field $a_\mu(x)$ transform as
\begin{equation}
\Psi'(x)= e^{-i\Lambda(x)} \Psi(x), \qquad a'_\mu(x)= a_\mu(x)+\partial_\mu \Lambda(x)
\end{equation}
where $\Lambda(x)$ is a (smooth) gauge transformation.

The second symmetry is invariance under the coordinate transformation of global rotations in real space,
\begin{equation}
x'_i=R_{ij}(\varphi) x_j
\end{equation}
where $R_{ij}(\varphi)$ is the $2 \times 2$ rotation matrix by an angle of $\varphi$. The Fermi field is invariant (a scalar) under rotations, $\Psi(Rx')=\Psi'(x)$. However the invariance of the action of Eq.\eqref{sec1sub2:effective} under global rotations requires that the Hubbard-Stratonovich fields ${\bm M}$, which are conjugate to the nematic order parameter field $Q_{ij}$ of Eq.\eqref{sec1sub2:quadrupolar}, transform not as a vector under rotations but as a {\em director}, i.e. a vector in without a direction. This means that it transforms under a rotation by {\em twice} the rotation angle in real space,
\begin{equation}
M'_i=R_{ij}(2\varphi)M_j
\label{eq:2phi}
\end{equation}
Under this transformation, the Hubbard-Stratonovich field is invariant under a rotation by $\pi$. Similarly, the nematic order parameter, i.e. the traceless symmetric $2 \times 2$ matrix field of Eq.\eqref{sec1sub2:quadrupolar}, transforms as a tensor under rotations by an angle of $2\varphi$.

\section{Effective Field Theory Of Nematic Order Parameter}
\label{LGNematic}

The full action of Eq.\eqref{sec1sub2:effective} is a quadratic form in the composite fermions. These fermionic fields can be integrated out  allowing us to  obtain an effective field theory for the nematic order parameter $M_1$ and $M_2$ coupled to the gauge fields. This procedure is safe provided on is expending about a saddle point state with a finite energy gap. 
The resulting effective Lagrangian can be decomposed as the three parts 
\begin{equation}
\mathcal{L}=\mathcal{L}_{a}+\mathcal{L}_{M}+\mathcal{L}_{a,M}
\label{eq:Leff-parts}
\end{equation}
 where  $\mathcal{L}_{a}$ and $\mathcal{L}_{M}$ include only the fluctuating gauge fields $\delta a + \delta A$ and only the nematic order parameter $M_{i}, i =1,2$, and $\mathcal{L}_{a, M}$ represents the coupling between the gauge fields and the nematic order parameter. 

For  clarity, we discuss the three parts, $\mathcal{L}_{a}$, $\mathcal{L}_{M}$, and $\mathcal{L}_{a,M}$, of the full effective theory separately. Here, we briefly show what we can learn from the three parts before describing  the details of each term. $\mathcal{L}_{a}$ is  the effective Lagrangian for the statistical gauge fields of the isotropic FQH states \cite{Lopez-1991}.   $\mathcal{L}_M$ is the effective Lagrangian for the nematic order parameters. It has the conventional Landau-Ginzburg form
 supplemented by a topological Berry phase term. We demonstrate that there is a continuous phase transition if the quadrupolar interaction $F_2$ is bigger than a critical value. Furthermore, we show that there is a Berry phase term for the nematic order parameter, which is similar to the Hall viscosity term, and the Berry phase term makes the quantum critical point have the dynamical exponent $z=2$. From ${\mathcal L}_{a,M}$, we will see that there is a {\it topological term}, similar to the Wen-Zee term, \cite{Wen_Zee} which describes the response to the curvature induced by {\it disclination} (not the deformation from the background geometry). In addition, $\mathcal{L}_{a,M}$ also contains an anisotropic Maxwell term that represents the coupling of the Kohn collective mode to the nematic order parameter fields \cite{you}.

\subsection{Gauge Field Lagrangian: ${\cal L}_a$}
\label{sec:La}

Here we consider the term of the effective Lagrangian of Eq.\eqref{eq:Leff-parts}  that includes only the gauge fields $a_{\mu}$ and $\delta A_{\mu}$. This part of the effective action does not know the nematic order parameter, and so it should be the same effective action of the gauge fields as in the isotropic FQH states \cite{Lopez-1991} 
\begin{align}
\mathcal{L}_a = -\frac{1}{2}(\delta a_{\mu} + \delta A_{\mu})\Pi^{0}_{\mu\nu}(\delta a_{\nu} + \delta A_{\nu}) + \frac{\varepsilon^{\mu\nu\lambda}}{8\pi} \delta a_{\mu}\partial_{\nu} \delta a_{\lambda},
\label{eq:Leff-CF}
\end{align}
Here $\Pi^{0}_{\mu\nu}(x-y)$, given by
\begin{equation}
\Pi^0_{\mu \nu}(x-y)=-i \frac{1}{Z_F} \frac{\delta^2 Z_F}{\delta a_\mu(x) a_\nu(y)}=\langle j_\mu(x) j_\nu(y)\rangle,
\label{eq:Pimunu0-def}
\end{equation}
 is the bare polarization tensor of the integer quantum Hall state of the composite fermion, and it is given in Ref.\cite{Lopez-1991} whose results we use. The current-current time-ordered correlators shown in Eq.\eqref{eq:Pimunu0-def} are computed in the free composite fermion theory and $Z_F$ is the partition function of composite fermions with an integer number of filled effective Landau levels. In the low energy and long wavelength limit $\Pi^0_{\mu \nu}({\bm q},\omega)$ is given by
\begin{align}
\Pi^0_{00}({\bm q},\omega)=&-\frac{1}{\pi} q^2\frac{m_e}{ {\bar b(1+\alpha\bar \omega^2_c)}},\nonumber \\
\Pi^0_{0j}({\bm q},\omega)=&-\frac{1}{\pi} q_j\omega \frac{m_e}{ {\bar b}(1+\alpha\bar \omega^2_c)}+ \frac{i}{\pi}\epsilon^{jk}q_k ,\nonumber\\
\Pi^0_{j0}({\bm q},\omega)=&-\frac{1}{\pi} q_j\omega \frac{m_e}{{\bar b}(1+\alpha\bar \omega^2_c)}-\frac{i}{\pi}\epsilon^{jk}q_k , \nonumber\\
\Pi^0_{ij}({\bm q},\omega)=&-\frac{1}{\pi} \delta_{ij}\omega^2\frac{m_e}{ {\bar b}(1+\alpha\bar \omega^2_c)} 
                                           -\frac{i}{\pi}\epsilon^{ij} \omega
                                           \nonumber\\
                                           & -\frac{(q^2\delta_{ij}-q_iq_j)}{ m_e (1+\alpha\bar \omega^2_c)}
\end{align}
where ${\bar b}=B/3$ for the $\nu=1/3$ Laughlin state. Here we have included in the results of Ref.\cite{Lopez-1991} the corrections due to the extra term in the kinetic energy of Eq.\eqref{eq:S4}.

\subsection{Order Parameter Lagrangian: $\mathcal{L}_M$}
\label{sec:LM}

We next obtain the Lagrangian for the nematic fields, $\mathcal{L}_M$, of Eq.\eqref{eq:Leff-parts}
 to the quartic order in the nematic order parameter by calculating one-loop Feynman diagrams. We will see that $\mathcal{L}_{M}$ exhibits the isotropic-anisotropic phase transition and the quantum phase transition. To calculate $\mathcal{L}_{M}$, we need to compute the two-point and four-point correlators of $N_{i} = -i\frac{\delta Z}{\delta M_i}$ and the calculation is done in Appendix \ref{app:calculation}.  In the main text, for simplicity we discuss only the case of the FQH state at the filling $\frac{1}{3}$.  However, as discussed in the Appendix \ref{app:calculation}, it is straightforward to generalize the calculations to the other states in Jain sequence $\nu = \frac{p}{2sp +1}, p,s \in {\mathbb Z}$. 

Integrating out the composite fermion and expanding about the low-energy limit, {\it i.e.,} taking the lowest terms in frequency $\omega$ and momentum $\bm{q}$, we obtain the effective theory of the nematic fluctuations 
\begin{align}
\mathcal{L}_{M}=& \frac{\epsilon^{ij}\bar{\rho}}{2(1+4\alpha\bar \omega^2_c)^2}  M_i \partial_0 M_j-r {\bm M}^2\nonumber\\
 -&\frac{\bar{\kappa}}{2} (\bm{\nabla}M_i)^2-\frac{u}{4}({\bm M}^2)^2. 
\label{nematic}
\end{align}
where ${\bm M}^2=M_1^2+M_2^2$. 

In the effective Lagrangian of Eq.\eqref{nematic} we have ignored two physically significant corrections terms. The Lagrangian of  Eq.\eqref{nematic} is invariant under  the $O(2)$ symmetry of arbitrary global rotations in the {\em order parameter space}, i.e. 
\begin{equation}
M_i \to R_{ij}(\phi) M_j,
\end{equation}
 where $R_{ij}(\phi)$ is the $2 \times 2$ rotation matrix by an arbitrary angle  $\phi$. However, as we saw in Section \ref{sec:symmetries}, the only symmetry (aside from gauge invariance) is a combination of a rotation {\em in space} by an angle $\varphi$ and a rotation in the {\em order parameter space} by $2 \varphi$, which leave  ${\bm M} \equiv -{\bm M}$ invariant. This means that the larger symmetry of the Lagrangian of Eq.\eqref{nematic} is only approximate and that the Lagrangian must contain terms which reduce the symmetry accordingly.  In fact,  the effective Lagrangian  allows for an extra  (formally irrelevant) operator of the form
\begin{equation}
\mathcal{L}_{SO}=-\lambda \Big( \big({\bm M} \cdot {\bm \nabla}\big) {\bm M} \Big)^2
\label{eq:SO}
\end{equation}
which is invariant under {\em joint} rotations in real space and in the order parameter space (and is formally a ``spin-orbit'' type coupling). Such terms are well known to arise in the free energy of classical liquid crystals \cite{deGennes-1993,chaikin-1995}.

The resulting effective Lagrangian of Eq.\eqref{nematic} has the same form as  the effective theory of the nematic order parameter in a Chern insulator\cite{you}, and of the effective field theory of the nematic FQH state of Maciejko and collaborators \cite{Maciejko2013}  Moreover, upon defining the complex field $\Phi=M_1+iM_2$, it is easy to see that the lagrangian of Eq.\eqref{nematic} is equivalent to the Lagrangian of a 2D dilute Bose gas (with $r$ playing the role of the chemical potential and $u$  the contact interaction). As in the Refs. \cite{you} and \cite{ Maciejko2013}, the effective  theory of the nematic order parameter field contains a Berry phase term associated with the non-dissipative response of the quantum Hall effect, which related to the Hall viscosity. This term makes time and space scale differently, and the associated quantum critical point has the dynamical exponent $z=2$. 

In our discussion we have neglected the role of the symmetries of the underlying lattice. While lattice  effects are irrelevant (and unimportant) for the topological properties of the FQH fluids they do matter for  the nematic fluctuations and ordering. In the case of  GaAs-AlAs heterostructures, the 2DEG resides on surfaces  which have a  tetragonal $C_4$ symmetry. The extra  terms of the Lagrangian that break the symmetry from the full continuous rotations down to $C_4$ are proportional to $M_1^2-M_2^2$ and $2 M_1M_2$. We will discuss in another section that these terms gap out the Goldstone modes of the nematic phase. 

The parameters entering into the the effective Lagrangian of Eq.\eqref{nematic} are obtained by a direct calculation of the correlators, and  are found to be
\begin{align}
r=&-\frac{1}{4F_2m^2_e}-\frac{{\bar \omega_c}}{2\pi {\bar l_b}^2 (1+4\alpha\bar \omega^2_c)}, \nonumber\\
{\bar \kappa}=&-\frac{\kappa}{2F_2m_e^2}\nonumber\\
&-\frac{1}{\pi} \Big[\frac{1}{(1+\alpha \bar \omega^2_c)}+\frac{1}{2(1+4\alpha \bar \omega^2_c)}+\frac{2}{(1+9\alpha \bar \omega^2_c)}\Big], \nonumber\\
u=&\frac{\bar b \bar\omega_c}{4\pi}\frac{1}{ (1+4\alpha \bar \omega^2_c)^2}\Big[\frac{1}{4 (1+4\alpha \bar \omega^2_c)}-\frac{3}{4(1+16\alpha \bar \omega^2_c)}\Big]
\label{nematic-parameters}
\end{align}
Here ${\bar \omega_{c}} = {\bar b}/m_e$ and ${\bar l_b}=\sqrt{3} \ell_0$ (for the Laughlin state at $\nu=1/3$) are the effective cyclotron frequency and the effective magnetic length of the composite fermion, where $\ell_0=B^{-1/2}$ is the magnetic length.

From these results we can also see that the nematic order parameter will condense only when the quadrupolar interaction is attractive and larger in magnitude  than the critical value 
\begin{equation}
|F_2^c| = \frac{\pi \bar{l_b}^2}{2 \bar{\omega_c} m^2_e}(1+4\alpha \bar \omega^2_c)
\label{eq:F2c}
\end{equation}
  Furthermore, since $u>0$ the quantum phase transition is continuous and the nematic state is stable.

From Eq.\eqref{sec1sub2:quadrupolar} it is clear that 
the nematic order parameters formally couple to the quadrupole density in the same way as the background metric couples to the energy-momentum tensor (although the extra term in the kinetic energy of Eq.\eqref{eq:S4} does not couple to the nematic fields). We can regard the nematic order parameters as a ``dynamical spatial metric" which modifies spatial components of the metric tensor. From this observation one may naively expect that the prefactor Berry phase term in Eq.\eqref{nematic} {\it may be} the Hall viscosity of the FQHE $\eta_{H}=\frac{{\bar \rho}}{2\nu}$ when $\alpha=0$. 

However, for $\alpha=0$, the prefactor of the Berry phase term of Eq.\eqref{nematic}
 is  the Hall viscosity term of the {\em integer quantum Hall state} 
at $\nu =1$, and not of the actual Hall viscosity of the {\em fractional} quantum Hall state. See the discussion of Sec. \ref{geometric-FQH}.
 This difference originates in the fact that the  ``dynamical metric'' associated to  the nematicity and the background metric {\em are not equivalent}. For FQH states, the nematic order parameters only couple with the stress energy tensor, while the background metric, not only couples with the stress energy tensor, but also appears in the form of a spin connection, as discussed in detail in Ref.\cite{Cho-2014}. In the composite fermion or composite boson theories, when we attach flux to the electron to form a composite particle, each flux quantum attached to the particle induces the additional angular momentum $1/2$. This makes the composite particle couple to the spin connection though the particle  is a scalar and not a spinor. The orbital spin then couples to the local geometry to the spin connection much  in the same way as  relativistic fermions do. Thus, after we perform flux attachment to describe the FQH fluids, the composite fermion resulting from the flux attachment will minimally couple with the spin connection $\omega_{\mu}$, as shown explicitly in Ref.\cite{Cho-2014}. The coupling through the spin connection with the background geometry is the origin of the difference between the nematic order parameter and the (deformed) background metric. The derivation of the correct Hall viscosity from the background metric deformation through the composite fermion theory was reported elsewhere \cite{Cho-2014}.

The results of this section can be easily generalized to all the states in the Jain sequence $\nu=\frac{p}{2p+1}$, with the effective Lagrangian density. However, when $p$ goes to infinity, the theory approaches to the half-filled Landau level and the gap vanishes. In this regime the system becomes  a non-Fermi liquid and the effective Lagrangian for the gauge field given by Eq.\eqref{eq:Leff-CF} now has  a Landau damping term \cite{halperin-1993,Polchinski-1994}. In this limit, at least formally, this theory is a generalization of the theory of the nematic quantum phase transition in Fermi fluids \cite{oganesyan-2001} to describe the compressible nematic quantum fluid at half-filled Landau levels (see Ref.\cite{Fradkin-2010} and references therein.)

\subsection{Order Parameter and Gauge Field Lagrangian: $\mathcal{L}_{a,M}$}
\label{emcoupling-nematic}

Here we derive the third term of the effective Lagrangian of Eq.\eqref{eq:Leff-parts},
$\mathcal{L}_{a,M}$, that describes the coupling between the gauge fields and the nematic order parameters. This part of the effective Lagrangian  will be important later for investigating the quantum numbers and statistics of the disclinations of the nematic phase. 

In the presence of  nematic order, the natural coupling between the nematic order parameter and the gauge field is as a local anisotropy of the Maxwell term. Since the order parameter acts as the spatial components of a metric tensor \cite{you}, the indices of the field strength tensor $f_{ij}$ of the gauge fields contract with the (inverse of) metric spatial tensor$g^{ij}$. The resulting terms in the effective Lagrangian are
\begin{align}
\mathcal{L}_{a,m}=&\frac{m_e 2M_1}{4\pi \bar{b}(1+4\alpha \bar \omega^2_c)} (\partial_x\delta \tilde{A}_0- \partial_0 \delta \tilde{A}_x)^2\nonumber\\
-&\frac{m_e 2M_1}{4\pi \bar{b}(1+4\alpha \bar \omega^2_c)} (\partial_y \delta \tilde{A}_0- \partial_0 \delta \tilde{A}_y)^2\nonumber\\
+&\frac{m_e M_2}{\pi \bar{b}(1+4\alpha \bar \omega^2_c)} (\partial_x \delta \tilde{A}_0- \partial_0 \delta \tilde{A}_x)(\partial_y \delta \tilde{A}_0- \partial_0 \delta \tilde{A}_y)
\end{align}
where $\delta \tilde{A}=\delta A+\delta a$.
These terms are second order in derivatives and are time-reversal and parity invariant.

However, there are contributions to  $\mathcal{L}_{a,M}$ which are first order in derivatives and hence break time-reversal and parity. These contributions have the form of a Wen-Zee term \cite{Wen_Zee,Hoyos2012}. The Wen-Zee term can be understood as the response of the FQH states to a change of the geometric curvature: the curvature will trap the gauge charge.  While this term can be ignored if the nematic order is uniform in space, it  has  interesting consequences for the charge and the statistics of the disclination in the nematic phase.

 To obtain the Wen-Zee term for the nematic order parameter we perform calculation of  one-loop diagrams with one current and one and two nematic fields,
\begin{align}
\mathcal{L}_{wz}=&-\frac{1}{2}T_{i \mu}M_i (\delta a_{\mu}+\delta A_{\mu}+2Z_{\mu})\nonumber\\
&+\frac{1}{3}R_{ij\mu}M_i M_j (\delta a_{\mu}+\delta A_{\mu}),
\end{align}
where we $T_{i\mu}$ and $R_{ij\mu}$ denote the following three-point (time-ordered) correlators of the composite fermions
\begin{align}
T_{i \mu} (\bm{r}, t)=&i2\frac{1}{Z_F}\frac{\delta Z_F}{\delta M_i \delta a_{\mu}} = -i \langle N_i (\bm{r}, t) j_{\mu} (0,0) \rangle,\nonumber\\
R_{ij\mu}[{\bm r}_i,t_i]=&-3 i\frac{1}{Z_F}\frac{\delta Z_F}{\delta M_i \delta M_j \delta a_{\mu}}\nonumber\\
=&  -\langle N_i (\bm{r}_1,t_1) N_j (\bm{r},t_2) j_{\mu} (\bm{r}_3, t_3) \rangle,
\label{eq:wz-correlators}
\end{align}
where the correlators are time-ordered functions of the free composite fermion theory, and
\begin{align}
Z_0=&0\nonumber\\
Z_x= &(\delta a_{x}+\delta A_{x})M_1+(\delta a_{y}+\delta A_{y})M_2\nonumber\\
Z_y= &(\delta a_{x}+\delta A_{x})M_2-(\delta a_{y}+\delta A_{y})M_1,
\end{align}
Diagrammatically the correlators of Eq.\eqref{eq:wz-correlators} are represented by the Feynman diagrams of  Fig.\ref{fig:wenzee}. They are computed explicitly in Appendix \ref{app:mixed-correlators}.

After calculating the above correlators, we obtain the coupling between the geometric curvature induced by the nematic fields and the statistical gauge field, which explicitly has the form of a Wen-Zee term
\begin{equation}
\label{eq:wenzee}
\mathcal{L}_{wz}=\frac{1}{4\pi} \epsilon^{\mu \nu \rho}\omega^Q_{\mu} \partial_{\nu} (\delta a_{\rho} +\delta A_{\rho})
\end{equation}
where $\omega^Q_\mu$ (with $\mu=0,x,y$) is the the effective spin connection induced by the local nematic order parameters, i.e.
\begin{align}
\omega^Q_{0}=&\frac{\epsilon^{ij}}{(1+4\alpha \bar \omega^2_c)^2} M_i \partial_0 M_j, \nonumber\\
\omega^Q_{x}=&\frac{\epsilon^{ij}}{(1+4\alpha \bar \omega^2_c)^2} M_i \partial_x M_j-t(\partial_x M_2-\partial_y M_1), \nonumber\\
\omega^Q_{y}=& \frac{\epsilon^{ij}}{(1+4\alpha \bar \omega^2_c)^2} M_i \partial_y M_j +t(\partial_x M_1+\partial_y M_2),
\end{align}
where
\begin{equation}
t=\frac{2}{1+\alpha \bar \omega^2_c} -\frac{2}{2+8\alpha \bar \omega^2_c} 
\end{equation}
The spin connection  $\omega^Q$  of the nematic order parameter is {\it different} than the spin connection of the background geometry. The meaning of the spin connection can be clarified by looking at its curl, 
\begin{equation}
\partial_x \omega^Q_y-\partial_y \omega^Q_x\propto \frac{1}{2}\sqrt{g} R
\end{equation}
where $R$ is the geometric curvature of the dynamical metric induced by the nematic order parameters $M_i$. Here $g$, the determinant of the metric, is given by $g=1-4 {\bm M}^2$.

Here the coupling term between the ``spin connection" and gauge fields in Eq.\eqref{eq:wenzee} has a similar form of the Wen-Zee term of Ref.\cite{Wen_Zee}. However, the coefficient in  Eq.\eqref{eq:wenzee} {\em is not the orbital spin} of the FQH state.
Instead this coefficient is equal to the orbital spin of the integer quantum Hall state at $\nu = 1$(when  $\alpha=0$). This can be easily understood from the composite fermion theory because the composite fermions  effectively are an the integer quantum Hall state and any response at the mean-field approximation of the composite fermion will be the same as that of the integer quantum Hall phase. This fact still remains true even after integrating out the statistical gauge field. The difference again comes from the nonequivalence between the nematic order parameter and the background metric. When we attach Chern-Simons flux to the fermion in a background metric, the orbital spin induced by the flux attachment also gives rise to additional geometry-gauge coupling term which has the form of a Wen-Zee term \cite{Cho-2014}. For the nematic order parameter, the coupling between the gauge field and the nematic order parameters only comes from the composite fermion which forms an IQHE. The derivation of the correct Wen-Zee terms through the composite fermion theory is reported in Ref.\cite{Cho-2014}.

\subsection{Full Effective Action} 
\label{sec:full}

Here we now ready to present the full effective Lagrangian of Eq.\eqref{eq:Leff-parts} in terms of the gauge fields and nematic order parameters. It is given by 
\begin{align}
\mathcal{L}=& \frac{\bar{\rho}\epsilon^{ij}}{2(1+4\alpha \bar \omega^2_c)^2}  M_i \partial_0 M_j-r {\bm M}^2 \nonumber\\
-&\frac{\bar{\kappa}}{2} (\bm{\nabla}M_i)^2
-\frac{u}{4}({\bm M}^2)^2\nonumber \\
 +& \frac{1}{4\pi} \epsilon^{\mu \nu \rho}\omega^Q_{\mu} \partial_{\nu} (\delta a_{\rho} +\delta A_{\rho})\nonumber\\
-&\frac{1}{2} \Pi^{0}_{\mu\nu}(\delta a_{\mu} + \delta A_{\mu})(\delta a_{\nu} + \delta A_{\nu}) \nonumber\\ 
+& \frac{1}{8\pi} \varepsilon^{\mu\nu\lambda} \delta a_{\mu}\partial_{\nu} \delta a_{\lambda} + \frac{1}{24\pi}\varepsilon^{\mu\nu\lambda}\omega^Q_{\mu} \partial_{\nu} \omega^{Q}_{\rho}
\label{FinalEffective}
\end{align}
In the last line we have added the gravitational Chern-Simons term of the induced spin connection of the nematic fields, where we used the results of Ref. \cite{Abanov2014}.

\section{Condensation of the GMP mode at the Nematic  Phase Transition}
\label{Nematic-QPT}

The FQH fluids have several  types of collective excitations \cite{gmp,Lopez-1993} The Kohn mode is a cyclotron collective mode related with the inter-Landau level excitation. If the system has  Galilean invariance, the energy of the Kohn mode at zero momentum only depends on the bare mass of the electron and is insensitive to any other microscopic detail \cite{kohn-1961} In a FQH fluid the Kohn mode is not the lowest energy collective excitation and at finite wave vector ${\bm q}$ can (and does) decay to lower energy modes. On the other hand, the lowest energy collective mode, the GMP mode, is stable. This mode is a quadrupolar intra-Landau level  fluctuation, and at long wavelengths it can be regard as a fluctuating quadrupole with structure factor $\sim {\bm q}^4$ (instead of ${\bm q}^2$ as in the case of the Kohn mode)  (see Refs.\cite{Lopez-1993,Yang-2012}). Therefore, for a FQH state with the quadrupolar interaction, we can expect that the interaction can change substantially  the behavior of the GMP mode by mixing with the nematic fluctuations (which are also quadrupolar). In this section, we consider the behavior of the GMP collective excitation of the FQH state at and near the quantum phase transition between the isotropic state and the nematic state.

To get the spectrum of the collective excitations, we need   the full polarization tensor of the electromagnetic response \cite{Lopez-1991}. To this end we first calculate the polarization tensor of the composite fermions
 \begin{equation}
\Pi^0_{\mu\nu}=-i\frac{1}{Z_F}\frac{\delta^2 Z_F}{\delta a_\mu \delta a_\nu},
\end{equation}
 where $Z_F$ is the partition function of the composite fermions. Because the composite fermion system is in an integer quantum Hall ground state, the poles of $\Pi_{\mu\nu}$ correspond to the Landau levels spaced by ${\bar \omega}_{c}$, the effective cyclotron frequency of the composite fermions (modified by the contributions of the extra terms of Eq.\eqref{eq:S4}). 
 
 Next we compute the change in $\Pi_{\mu \nu}$ due to the effects of both the quadrupolar interaction and of the density-density interaction and determine the full polarization tensor for the external electromagnetic field $K_{\mu \nu}$. The current and the nematic fields are defined in terms  of the composite fermion $\Psi$ by
\begin{align}
j_{i} = \frac{\delta S}{\delta a_\mu}, \quad N_{i} =  \frac{\delta S}{\delta M_i} = \Psi^{\dagger} T_{i} \Psi. 
\end{align}
where $S$ is the full action of Eq.\eqref{sec1sub2:effective} supplemented by the additional term of Eq.\eqref{eq:S4}.
 
We next compute the current-current correlators including the mixing with the nematic fields to lowest orders in the quadrupolar coupling $F_2$.
To this end  we first  calculate the polarization tensor $\Pi_{\mu \nu}$   to include de effects of the nematic fluctuations to lowest order in the quadrupolar interaction $F_2$. 
This calculation involves summing over all one-particle-reducible diagrams, i.e. an infinite series of bubble diagrams with two external gauge fields and arbitrary number of  quadrupolar insertions connecting the bubbles pairwise. The result of this RPA-type computation 
is
\begin{align}
\Pi_{ij}({\bm q},\omega)=& \Pi^{0}_{ij}+2F_2 m_e^2\sum_{a,b}\langle  j_i N_a \rangle  \langle  N_b j_j \rangle \nonumber\\
+& (2F_2m_e^2)^2\sum_{a,b}\langle  j_i N_a \rangle \langle  N_a N_b\rangle \langle  N_b j_j \rangle +\cdots \nonumber\\
=& \Pi^{0}_{ij}+\frac{2F_2m_e^2 \sum_{a,b} \langle  j_i N_a \rangle \langle  N_b j_j \rangle}{1-(2F_2m_e^2) \sum_{a,b}\langle  N_a N_b \rangle}
\label{eq:Pimunu-RPA}
\end{align}
(where we have set $\kappa=0$).
Here $\Pi^{0}_{ij}$ is the polarization tensor for the statistical gauge field of the composite fermions with $\nu =1$, $\langle  N_a N_b \rangle $ 
is the correlator matrix of the nematic order parameters, and $\langle j_\mu N_a \rangle $ is the mixed correlator of a current and a nematic field, 
both of which were  calculated in the previous section. 
To simplify the notation, in Eq.\eqref{eq:Pimunu-RPA} we dropped the explicit momentum and frequency dependence of the correlators.

Since we are interested in the low energy and long wave-length limit, we expand $\langle  N_a N_b \rangle$ in the leading order for both the momentum 
$\bm q$ and the frequency $\omega$ and obtain
\begin{align}
\langle  N_1 N_1 \rangle=& \langle  N_2 N_2 \rangle =2i \frac{1}{Z_F}\frac{\delta Z_F}{\delta M_1 \delta M_1}\nonumber\\
=&\frac{4 {\bar \omega_c}^3}{{\bar l_b}^2 \pi (\omega^2-4{\bar \omega_c}^2)(1+4\alpha \bar \omega^2_c )^2},\nonumber\\
\langle  N_1 N_2 \rangle =&- \langle  N_2 N_1 \rangle=2i \frac{1}{Z_F}\frac{\delta Z_F}{\delta M_1 \delta M_2}\nonumber\\
=&\frac{2 \omega {\bar \omega_c}^2}{i \pi {\bar l_b}^2 (\omega^2-4{\bar \omega_c}^2)(1+4\alpha\bar \omega^2_c)^2 },
\end{align}

Using the results for the polarization tensor $\Pi_{ij}$ of Eq.\eqref{eq:Pimunu-RPA} 
we find the following effective Lagrangian for the gauge fields 
\begin{align}
\mathcal{L}_{a}=&-\frac{1}{2} \Pi_{\mu \nu}(\delta a_{\mu}+\delta A_{\mu}) (\delta a_{\nu}+\delta A_{\nu})\nonumber\\
&+ \frac{1}{8\pi} \epsilon^{\mu \nu \rho} \delta a_{\mu}\partial_{\nu} \delta a_{\rho}
-\int d^{2}x' \frac{V(x-x')}{32\pi^{2}} \delta b(x) \delta b(x')
\end{align}
 Integrating out the statistical gauge field $\delta a_{\mu}$, we finally obtain the full  response function $K_{\mu \nu}$ for the external electromagnetic fields
\begin{align}
K_{00}=&q^2 K_0, \nonumber \\
K_{0i}=&\omega q_i K_0+i \epsilon_{ik} q_k K_1, \nonumber\\
K_{i0}=&\omega q_i K_0-i \epsilon_{ik} q_k K_1, \nonumber\\
K_{ij}=&\omega^2 \delta_{ij} K_0-i\epsilon_{ij} \omega K_1+(q^2 \delta_{ij}-q_iq_j)K_2, \nonumber\\
\mathcal{L}_{A}=&K_{\mu \nu} \delta A_{\mu} \delta A_{\nu},
\end{align}
where  $K_{\mu}$ is  given by 
\begin{align}
K_0=&-\frac{\Pi_0}{16\pi^2 D} \nonumber\\
K_1=&\frac{1}{4\pi }+\frac{\Pi_1+\frac{1}{4\pi}}{16\pi^2 D}+\frac{V({\bm q})\Pi_0 {\bm q}^2}{64\pi^3 D} \nonumber\\
K_2=&\frac{\Pi_2}{16\pi^2 D}+\frac{V({\bm q})(\omega^2\Pi_0^2-\Pi_1^2)}{ D}+\frac{V(q)\Pi_0 \Pi_2 {\bm q}^2}{ D} 
\label{eq:Kmunu}
\end{align}
and $D$ is
\begin{equation}
D=\Pi_0^2 \omega^2-(\Pi_1+\frac{1}{4\pi })^2+\Pi_0(\Pi_2-\frac{V({\bm q})}{16\pi^2}){\bm q}^2
\end{equation}
In the above expressions $\Pi_i$ are frequency and momentum-dependent  functions whose explicit form can be found  in Ref.[{\onlinecite{Lopez-1991}.

The poles in $K_{\mu \nu}$ give the spectrum of the collective excitations of the FQHE of Ref.\cite{Lopez-1991}, generalized to include
both the quadrupolar and the density-density interactions.
At long wavelengths, this correlator has a pole at $3\bar\omega_{c}$ (with residue $\sim {\bm q}^2$), the cyclotron frequency of the 
{\it electron} (recall that $\bar\omega_c$ is the effective cyclotron frequency of the composite fermion). 
This pole is identified as the (cyclotron resonance) Kohn mode \cite{kohn-1961}, slightly shifted here by the extra term we added to the kinetic energy 
(Eq.\eqref{eq:S4}).

On the other hand, we find that the attractive quadrupolar interaction pushes down to lower energies the lowest collective excitation, the Girvin-MacDonald-Platzman (GMP) mode \cite{gmp} (which has residue $\sim {\bm q}^4$). This mode has the dispersion 
\begin{equation}
\omega^2=\omega^2_{1}+(\alpha_{1,2}{\bar \omega_c}^3 -\frac{F_2m^2_e {\bar \omega_c}^3 \kappa}{{\bar l_b}^4}) (q{\bar l_b})^2
\end{equation}
where we have set
\begin{align}
\omega_{1}=&\frac{4 \tilde{F_2} }{\pi}+ 2 \bar \omega_c(1+4\alpha\bar \omega^2_c ),\nonumber\\
\alpha_1=&\frac{\omega^2_1-\bar \omega'^2_c}{\bar \omega'^2_c-\bar \omega'_c \omega_1+2(\omega_1^2-\bar \omega'^2_c)} \frac{1}{(c_1\omega_1-c_2)t^2},\nonumber\\
\alpha_2=&- \frac{\omega^2_1-\bar \omega'^2_c}{\bar \omega'^2_c+\bar \omega'_c \omega_1+2(\omega_1^2-\bar \omega'^2_c)} \frac{1}{(c_1\omega_1+c_2)t^2}
\end{align}
and where we used the notation
\begin{align}
c_1=&\frac{-\tilde{F_2} \bar \omega_d}{2(4\bar\omega_d^2-\omega_1^2)}\left(1-\frac{8\tilde{F_2}\bar\omega_d}{\pi(\omega_1^2-\bar\omega_d^2)}\right),\nonumber\\
c_2=&\frac{\tilde{F_2}^2 \bar\omega_d \omega_1^2}{(4\bar\omega_d^2-\omega_1^2)^2},\nonumber\\
\tilde{F_2}=&\frac{F_2m^2_e \bar\omega_c^2}{\bar l_b^2},\nonumber\\
\bar \omega'_c=&\bar \omega_c(1+\alpha \bar \omega^2_c),\nonumber\\
\bar \omega_d=&\bar \omega_c(1+4 \alpha \bar \omega^2_c)\nonumber\\
t=&\frac{2}{1+\alpha \bar \omega^2_c}-\frac{2}{2+8\alpha \bar \omega^2_c}.
\end{align}

It is easy to check that at the nematic transition of Eq.\eqref{nematic}, where the ``nematic mass'' $r\to 0$ at the critical value of the quadrupolar 
interaction $F_2^c$ (given in Eq.\eqref{eq:F2c}), the gap of the GMP mode vanishes, $\omega_1 \to 0$. It is also easy to see  that near the phase 
transition $\alpha_1<0$, $\alpha_2<0$ , and   $F_2 \kappa<0$. 
Now, provided $\alpha_{1,2} -\frac{F_2m^2_e \kappa}{\bar l_b^4} > 0$ near and at the transition, then the GMP mode will condense at zero momentum. 
This will result a nematic phase and the FQH fluid will spontaneously break the rotational symmetry. This condition can be achieved provided the range of 
the quadrupolar interaction, controlled by $\kappa$, is large enough. 
On the other hand, if $\alpha_{1,2} -\frac{F_2m^2_e \kappa}{\bar l_b^4}<0$, the GMP mode will condense at a finite momentum. This would result a 
crystalline phase in which electrons break spontaneously the translational and rotational symmetries of the two-dimensional plane. In both cases, the Kohn 
mode remains gapped at and near the transition, and thus the liquid crystalline phases are incompressible electronic liquid states and have the quantized 
Hall response. 

Early numerical results by Scarola, Park and Jain \cite{Scarola2000}  predicted that for certain type of interactions the FQH fluid would become 
unstable to an  uniform exciton condensate associated with the GMP mode. 
Our results show that their exciton condensate is equivalent to a 
quantum phase transition to a nematic state.

\section{Goldstone Mode and Disclinations in Nematic Phase}
\label{sec:disclinations}

We now discuss the properties of the nematic phase. There are several particle-like excitations in the phase. First of all, the Kohn mode and the Laughlin quasi-particles remain  massive. The only change is that their propagation is anisotropic.  In addition to of these excitations, there are two more excitations which are absent in the isotropic phase. 

\subsection{Goldstone Modes}

The nematic order parameter breaks the (continuous) rotational symmetry of two dimensional plane and thus there is an associated Goldstone mode and an amplitude mode (which is strongly mixed with the GMP mode). The spectrum of the nematic Goldstone mode can be obtained straightforwardly from $\mathcal{L}_M$.  In the low energy regime and deep enough in the nematic phase, $r=-|r |<0$, we can  consider the effective Lagrangian of the phase fluctuations (the Goldstone mode). Similarly to the effective Lagrangian in the Bogoliubov theory of superfluidity (or in the composite boson theory of the FQHE) in the nematic phase the amplitude fluctuations yield an effective Lagrangian for the Goldstone boson (the phase field $\theta$) of the form
\begin{equation}
\mathcal{L}_{M}=\frac{1}{2} \frac{\rho_n}{v_n}  (\partial_0 \theta)^2-\frac{1}{2} \rho_n v_n |{\bm \nabla} \theta|^2+....
\end{equation}
where the nematic stiffness $\rho_n$ and the velocity of the Goldstone modes $v_n$ are given by
\begin{equation}
\rho_n=\frac{\sqrt{|r | {\bar \kappa}}}{2u (1+4\alpha {\bar \omega}_c^2)}, \qquad
v_n=4 \sqrt{|r | {\bar \kappa}}(1+4\alpha {\bar \omega}_c^2)
\end{equation}
where the parameters $r$, ${\bar \kappa}$ and $u$ are given in Eq.\eqref{nematic-parameters}. In the nematic phase the Goldstone bosons are massless and have a linear dispersion
\begin{equation}
\omega(\bm q)=v_n |{\bm q}|
\label{eq:Goldstone-mode-dispersion}
\end{equation}

On the other hand, the nematic Goldstone mode will be gapped if there is a explicit weak symmetry breaking term. For instance, if the underlying lattice has tetragonal symmetry, the  point group symmetry of the 2DEG is be $C_4$. For the nematic order parameter, which is invariant under rotations by $\pi$,  this term reduces the symmetry to a $\mathbb{Z}_2$ (Ising) symmetry. The appropriate symmetry breaking term in the nematic Lagrangian has the form 
\begin{align}
\mathcal{L}_{SB}=&-\gamma_1 (M_1^2-M_2^2)-\gamma_2 2 M_1 M_2 \nonumber\\
=&-  \gamma_1 M^2 \cos 4\theta - \gamma_2 M^2 \sin 4 \theta
\end{align}
where $\gamma_1$ and $\gamma_2$ are two coupling constants.
 In this case, the mass gap of the Goldstone mode is linearly proportional to the strength of these weak symmetry breaking terms.  

\subsection{Disclinations}

In $D=2$ space dimensions nematic order parameters have topological singularities  (or defects ) called disclinations \cite{chaikin-1995}. Due to the presence of the goldstone mode, the disclinations experience logarithmic interaction between them. In 2D disclinations are half-vortices of the order parameter director field ${\bm M}$.  When two disclinations are separated by a distance $R$, the energy cost for the configuration is 
\begin{equation}
E=\int_{R>|{\bm x}|> l_0} d^2x\; \kappa {\bm M}^2 \frac{1}{{\bm x}^2}=\kappa {\bm M}^2 \ln (R/l_0), 
\end{equation}
where $l_{0}$ is a ultra-violet cut-off for the integral which can be taken to be the correlation length of the nematic order parameter in the nematic phase. Hence, at zero temperature disclinations and anti-disclinations are bound in (neutral) pairs but, above a critical temperature $T_c$, they proliferate 

We will now show that in the nematic FQH state, the disclination  carries electric charge due to the Wen-Zee coupling between the spin connection defined by the nematic fields $\omega^{Q}_{\mu}$ and the gauge fields in the effective action $\mathcal{L}$ Eq. \eqref{FinalEffective}. To investigate the charge accumulated at the disclination, we first integrate out the statistical gauge field $\delta a_{\mu}$ in the effective action to find a Wen-Zee term in the effective action 
\begin{equation}
\mathcal{L}_{\omega^{Q}, \delta A} = \frac{1}{12\pi} \epsilon^{\mu \nu \rho}\omega^Q_{\mu} \partial_{\nu}\delta A_{\rho}. 
\end{equation}
where $A_\mu$ just an external weak electromagnetic probe. 
This term,  will allow us to compute the electric charge of the disclination. Maciejko {\it etal.} \cite{Maciejko2013} use the term ``nematic gauge field'' to refer to what here we call the nematic spin connection, $\omega_\mu^Q$.

Let us consider the case in which there exists a $\pi$ nematic vortex centered at ${\bm x}=0$. (Recall that since the nematic fields are directors their orientation is defined mod $\pi$.) 
We can calculate the electric charge accumulated at the disclination. We find
\begin{align}
\delta \rho (\bm x) = \frac{1}{12\pi} (-\frac{{\bm M}^2\delta(\bm x)}{(1+4\alpha \bar \omega^2_c)^2}- t\frac{|{\bm M}| \cos2\theta}{{\bm x}^2})
\label{vortex}
\end{align}
The first term indicates the spin connection of the nematic disclination act as the gauge field of a single flux at the disclination core. Thus, a disclination of the nematic field serves as a particle source which changes the local charge density. The second term indicates the charge density gets redistributed  as a result of non-zero quadrupole moment. In classical electrodynamics, the non-uniform charge density could give rise to an electron quadrupole moment $Q_{ij}=\int d^2x \rho(r) (2x_i x_j-\delta_{ij} |{\bm x}|^2)$ and vice versa. Since our nematic field couples to the stress tensor, a nematic order with a disclination configuration  leads to a new charge density distribution,  shown in the second term in Eq.\eqref{vortex}. The charge of the disclination depends on the strength of the order parameter $|\bm M|$ and is not quantized. This implies that the disclinations will generally have irrational mutual statistics with quasiparticles and irrational self statistics. 
Most of these  results were anticipated on phenomenological and symmetry grounds in the work of Maciejko {\it et al.} \cite{Maciejko2013}.

\section{Response of the Nematic FQH fluid to changes in the geometry}
\label{geometric-FQH}

In this section, we would explore the  response of nematic FQH fluid to  a long wavelength change in the geometry of the underlying surface (i.e. the crystal) on which the 2DEG is defined such as a shear distortion. Changes in the geometry can be described in terms of a background spatial metric $g_{ij}$
\begin{align}
g_{ij}=
\begin{pmatrix} 
 1-2e_1  &- 2e_2\\ 
  -2e_2 &  1+2e_1
\end{pmatrix}
\end{align}
which modifies the form of the action of Eq.\eqref{sec1sub2:effective} to the following expression
\begin{align}
\mathcal{S} = \int &d^{2}x dt \left[ \Psi^{\dagger}(x) D_0 \Psi(x)-\frac{1}{2m_e}\left( {\bm D}\Psi (x) \right)^{\dagger} \cdot \left( {\bm D}\Psi(x) \right)\right] \nonumber\\
 -\int & d^{2}x dt ~V \frac{\delta b(x) ^2}{32\pi^{2}}  \nonumber\\
+ \int & d^{2}x dt \frac{\varepsilon^{\mu\nu\lambda}}{8\pi} \delta a_{\mu}\partial_{\nu} \delta a_{\lambda} \nonumber\\ 
+ \int & d^{2}x dt  \Big[ \frac{1}{4F_2 m^2_e} {\bm M}^{2}\nonumber\\
 +& \frac{\kappa}{4F_2 m^2_e}  \sum_{i=1,2} \bm{\nabla} M_{i} \cdot \bm{\nabla}M_{i} \nonumber\\ 
+&\frac{\kappa}{4F_2 m^2_e}  \Big(2e_1(\partial_x^2-\partial_y^2)+2e_2  2\partial_x\partial_y\Big)M^2_i\nonumber\\ 
+& \frac{M_{1}+e_1}{m_e} \Psi^{\dagger}\Big(D^2_x  - D^2_y\Big) \Psi\nonumber\\ 
+ &\frac{M_{2}+e_2}{m_e} \Psi^{\dagger} \Big(D_x D_y + D_y D_x\Big)\Psi \nonumber\\
 -& \alpha\Psi^{\dagger} \Big(-\frac{{\bm D}^2}{2m_e}-\frac{\bar\rho \pi}{m_e}\Big)^3 \Psi \nonumber\\
+&\frac{2 {\bm e} \cdot {\bm M}}{m_e}\Psi^{\dagger} {\bm D}^2\Psi\Big]
\end{align}
where the covariant derivative now is 
\begin{equation}
D_{\mu}=\partial_{\mu}+i(A_{\mu}+a_{\mu}+\omega^{b}_{\mu}),
\end{equation}
 where $\omega^{b}$ is the spin connection of the {\em background metric}. Here we have used the result from our recent work \cite{Cho-2014} that upon attaching two Chern-Simons flux to the fermion, the composite particle effectively carries spin $1$ and couples to the spin connection of the background geometry. Notice also that the quadrupolar interaction is modified by a change of the geometry. To simplify matters we considered only a contact density-density coupling (parametrized by the interaction strength $V$).

We can now make use of the results of the preceding sections (and of the results of Ref.\cite{Cho-2014}) to derive the effective Lagrangian for the nematic fields ${\bm M}$ at low energies and long distances. It is given by
\begin{align}
\mathcal{L}_{M}=&\epsilon^{ij} \frac{\bar{\rho}}{2}  \Big(\frac{M_i}{1+4\alpha \bar \omega^2_c}+e_i\Big) \partial_0 \Big(\frac{M_j}{1+4\alpha \bar \omega^2_c}+e_j\Big)
\nonumber\\
+&\frac{\bar\omega_c}{2 \bar \pi \bar l^2_b}|\bm e|^2 
+\epsilon^{ij} \bar{\rho} ~ e_i \partial_0 e_j \nonumber\\
+&\frac{1}{12\pi}\epsilon^{\mu \nu\lambda}\Big(A_\mu +\omega_\mu^b+\frac{1}{2}\omega_\mu^m\Big)\partial_\nu\Big(A_\lambda+\omega_\lambda^b+\frac{1}{2}\omega_\lambda^m\Big)\nonumber\\
-&r |\bm M|^2 -\frac{1}{48\pi}\epsilon^{\mu \nu \lambda}\omega_\mu^m \partial_\nu \omega_\lambda^m -\frac{u}{4}({\bm M}^2)^2
\end{align}
Here $\bar \rho$ is the average electron density,  and we denoted by $\omega^m$  the spin connection for the sum of the background and nematic metrics,
\begin{align}
\omega^m_{0}=&\epsilon^{ij}\Big(\frac{M_i}{1+4\alpha \bar \omega^2_c}+e_i\Big) \partial_0 \Big(\frac{M_j}{1+4\alpha \bar \omega^2_c}+e_j\Big) , \nonumber\\
\omega^m_{x}=&\epsilon^{ij}\Big(\frac{M_i}{1+4\alpha \bar \omega^2_c}+e_i\Big) \partial_x \Big(\frac{M_j}{1+4\alpha \bar \omega^2_c}+e_j\Big)\nonumber\\
                          &-\Big(\partial_x (tM_2+e_2)-\partial_y (tM_1+e_1)\Big), \nonumber\\
\omega^m_{y}=& \epsilon^{ij}\Big(\frac{M_i}{1+4\alpha \bar \omega^2_c}+e_i\Big) \partial_y \Big(\frac{M_j}{1+4\alpha \bar \omega^2_c}+e_j\Big)\nonumber\\
                         &+\Big(\partial_x (tM_1+e_1)+\partial_y (tM_2+e_2)\Big),
\end{align}
 To better illustrate the effect of nematic fluctuations, we  rewrite the action in terms of a separate  dependence on  the background metric and on the metric defined by the nematic order parameter, and obtain
\begin{widetext}
\begin{align}
\mathcal{L}_{M}=&\epsilon^{ij} \frac{\bar{\rho}}{2}  \Big(\frac{M_i}{1+4\alpha \bar \omega^2_c}+e_i\Big) \partial_0 \Big(\frac{M_j}{1+4\alpha \bar \omega^2_c}+e_j\Big)-
r |\bm M|^2 +\frac{\bar \omega_c}{2\pi \bar l_b^2}|\bm e|^2
+\epsilon^{ij} \bar{\rho} ~ e_i \partial_0 e_j-\frac{u}{4}({\bm M}^2)^2\nonumber\\
&+\frac{1}{12\pi}\epsilon^{\mu \nu\lambda}\Big(A_\mu+\frac{3}{2}\omega_\mu^b\Big)\partial_\nu\Big(A_\lambda+\frac{3}{2}\omega_\lambda^b\Big)-\frac{1}{12\pi}\epsilon^{\mu \nu\lambda} \omega_\mu^Q \partial_\nu A_\lambda-\frac{1}{48\pi} \epsilon^{\mu \nu\lambda}\omega_\mu^b \partial_\nu \omega_\lambda^b\nonumber\\
&+\frac{1}{12\pi(1+4\alpha \bar \omega^2_c)}\epsilon^{\mu\nu\rho}\Big(M_1\partial_{\mu}e_2-M_2\partial_{\mu} e_1+e_1\partial_{\mu} M_2-e_2\partial_{\mu}M_1\Big) \partial_{\nu} A_{\rho}\nonumber\\
&+\frac{1}{12\pi(1+4\alpha \bar \omega^2_c)}\epsilon^{\mu\nu\rho}\Big(M_1\partial_{\mu}e_2-M_2\partial_{\mu} e_1+e_1\partial_{\mu} M_2-e_2\partial_{\mu}M_1+(1+4\alpha \bar \omega^2_c)\omega^Q_{ \mu}\Big) \partial_{\nu} \omega^{b}_{\rho}
\label{nematiceffect}
\end{align}
\end{widetext}
In the isotropic phase, the nematic field is massive so it can be integrated out. This  generates operators which are higher order in derivatives and  are irrelevant in in the low energy and long distance regime of our theory. Finally, we obtain  the following simple expression for the effective theory of background metric in the symmetric phase,
\begin{align}
\mathcal{L}_{M}=&\epsilon^{ij} \frac{3\bar{\rho}}{2}  e_i \partial_0 e_j \nonumber\\
+&\frac{1}{12\pi}\epsilon^{\mu \nu \lambda} (A_\mu+\frac{3}{2} \omega_\mu^b)\partial_\nu (A_\lambda+\frac{3}{2}\omega_\lambda^b)\nonumber\\
-&\frac{1}{48\pi} \epsilon^{\mu \nu \lambda} \omega_\mu^b \partial_\nu \omega_\lambda^b
\end{align}
which is consistent with our recent results \cite{Cho-2014}.

In the nematic phase, the nematic order couples to the electrons (and the composite fermions) as an effect mass $m^{ab}$ tensor. In the isotropic phase, the Hall viscosity \cite{Avron1995,Tokatly-2009,Read2011,Bradlyn2012,Hoyos2012,son-2013,Cho-2014}, defined by $\eta^H=\eta^{xx}_{xy}=-\eta^{yy}_{xy}$, is isotropic and for the $\nu=1/3$ Laughlin FQH state it is found to be given by $\eta^H=\frac{3}{2} {\bar \rho}$, in agreement with earlier results.  In the nematic phase, provided the nematic order is uniform in space, the Hall viscosity $\eta^H=(\eta^{xx}_{xy}-\eta^{yy}_{xy})/2=\frac{3\bar{\rho}}{2}$ remains the same as in the isotropic FQH fluid phase. However, since the system is spatially anisotropic, we can define the  combination of the components of the viscosity $\eta^D=(\eta^{xx}_{xy}+\eta^{yy}_{xy})/2\propto \bar{\bm M}\eta^H$, which indicates that there is a  viscosity response for a mixed  shear  and a dilation deformation which, however, is not universal. 
In particular, when the nematic order is uniform in space, the geometry quantity such as the Hall viscosity, orbital spin, central charge remains unchanged at the universal value. 

On the other hand, in the presence of a disclination in the nematic phase, from Eq.\eqref{nematiceffect} we see that the Hall viscosity is modified. If ${\bm M}(\bm x)$ is a configuration of the nematic order parameter with a disclination at ${\bm x}={\bm x}_v$ with winding number $n_v$, the Hall viscosity of the fluid now is
\begin{align}
\eta(\bm x)=&\eta^H_0+\frac{1}{12\pi} \frac{|\bm M|^2}{(1+4\alpha \bar \omega^2_c)^2}  n_v \delta({\bm x}-{\bm x}_v) \nonumber\\
&+2t[(\partial^2_x-\partial^2_y)M_1+4\partial_x\partial_y M_2])
\end{align} 
The first term is equal to the Hall viscosity of the isotropic phase. The second term shows the change of the Hall viscosity due to the nematic disclination. Here, $n_v$ is the winding number of the disclination, and ${\bm x}_v$ is the coordinate of the disclination core. The third term indicates the charge density redistribution results from the nematic order as a quadrupole moment which affects the value of the Hall viscosity. However, the orbital spin and the gravitational Chern-Simons term remain the same in both phases.

In the recent work of Maciejko~ \cite{Maciejko2013}{\it et.al.} these authors  considered an effective description of the nematic FQH state and the transition between an isotropic state and the anisotropic state. They used a composite boson theory and wrote down the symmetry-allowed terms in the effective Lagrangian. Interestingly they  identified the coupling between the nematic order parameter and the statistical gauge field by making an analogy to the case of the magnetization of the quantum Hall ferromagnet. Furthermore, they found that the critical theory has the dynamical scaling exponent $z=2$ due to the Berry phase term for the nematic order parameter, which we also found here. However, in their description the coefficient of the Berry phase term is the full Hall viscosity of the FQH fluid. Instead, here we showed that  that the Berry phase term is not exactly equal to the Hall viscosity  of the FQH fluid but it is equal to  the Hall viscosity of  integer quantum Hall mean-field state of the composite fermions.

\section{conclusions}
\label{conclusions}

In this work, we have studied the nematic quantum phase transition inside a FQH state in a 2DEG in with an attractive  quadrupolar interaction between electrons. We used the Chern-Simons theory of composite fermions. Since the FQH state is gapped a critical attractive quadrupolar coupling is needed  for the system  to develop a finite quadrupole density and to break  rotational symmetry spontaneously. The quantum phase transition has dynamical exponent $z=2$ and it is in the universality class of the quantum phase transition in the dilute Bose gas. The $z=2$ quantum criticality is a consequence of a Berry phase term present in the effective action for the nematic fields  is related to (but not the same) as the Hall viscosity. We showed that the coefficient of the Berry phase term is the Hall viscosity of the mean field theory of the composite fermions in a  background nematic field. The actually Hall viscosity of the FQH fluid (both in the isotropic and in the nematic phase) is obtained as a response to a shear distortion of the geometry in which the electrons move.
Furthermore, we uncover the existence  of a geometric Chern-Simons term between the nematic order parameters and the gauge fields. The term is of the same form as Wen-Zee term and the ``spin connection'' is interpreted in terms of the order parameter instead of the background metric. Then the flux of the ``spin connection" is proportional to the disclination density in the nematic phase, and that as a consequence  the disclination carries non-quantized gauge charge and statistics. 

After the identification of the criticality and the phases, we investigated the   excitations near the quantum phase transition. As the the nematic quantum phase transition is approached,  the mass gap of GMP mode of the FQH fluid is shown to vanish continuously. On the other hand, the Laughlin quasiparticles and the Kohn mode remain gapped at the transition and thus the Kohn's theorem is not violated at and near the transition. Depending on the microscopic details of the interactions, we showed that the GMP mode can close its gap either at finite or at zero momentum, either giving rise to a nematic or to crystal (or stripe) phase. Both liquid crystalline phases obtained through the softening of the GMP mode are incompressible electronic liquid crystals and thus are expected to have a fractionally quantized Hall response. It is notable that the mechanism of the isotropic-nematic transition described here is special for FQH states. For the integer quantum Hall states, the lowest excitation is the Kohn mode, inter-Landau level excitation, and it cannot close its gap at zero momentum without a large amount of Landau level mixing and a strong violation of Galilean invariance.

\begin{acknowledgments}
We would like to thanks  Tankut  Can, Andrey Gromov, Taylor Hughes, Shamit Kachru, Steve Kivelson, Rob Leigh,  Joseph Maciejko, Roger Mong, Chetan Nayak, Kwon Park, Onkar Parrikar, Dam Thanh Son, Shivaji Sondhi, Paul Wiegmann, and Michael Zaletel  for helpful discussions. G.~Y.~C. thanks the financial support from ICMT postdoctoral fellowship. EF thanks the KITP (and the Simons Foundation) and its IRONIC14 program for support and hospitality. This work is supported in part by the by the National Science Foundation, under grants DMR-1064319 (GYC) and DMR 1408713 (YY,EF) at the University of Illinois, PHY11-25915 at KITP (EF), and by ICMT.
\end{acknowledgments}

\appendix

\section{Calculation of the Nematic Correlators}
\label{app:calculation}

In this appendix, we summarize the calculation of the correlators of the nematic order parameters which we extensively use in the main text. 

\begin{figure}[hbt]
   \centering
  \includegraphics[width=0.3\textwidth]{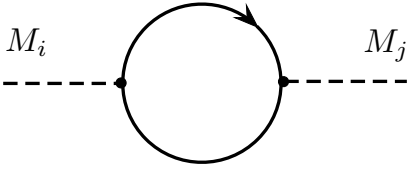}
\caption{Correlator of the nematic order parameters}
\end{figure}

First, we focus on the symmetric part of the correlators, which is the mass term for the order parameters. Here for simplicity, we set the magnetic length $\bar{l_b}$ and electron bare mass $m_e$ to be $1$. At the end, we restore the magnetic length and mass in the expression by dimensional analysis. For the filling fraction $\nu = \frac{p}{2p+1}$, the correlator is 
\begin{widetext}
\begin{align}
& \langle  N_1(\bm{r}_1, t_1) N_1(\bm{r}_2,t_2) \rangle=2i\frac{\delta Z}{\delta M_1 \delta M_1}
=-i\langle \mathcal{T} \Psi^{\dagger}(\bm{r}_1,t_1)(D^2_x -D^2_y)\Psi(\bm{r}_1,t_1)
\Psi^{\dagger}(\bm{r}_2,t_2)(D^2_x-D^2_y)\Psi(\bm{r}_2,t_2)\rangle \nonumber\\
&=-i\sum^{\infty}_{m>p}\sum^{p-1}_{l=0} \sum_{k_1,k_2} \bigg[ e^{i(\omega_m-\omega_l)(t_2-t_1)} \Theta(t_1-t_2) 
\phi_{l,k_1}^{\dagger}(\bm{r}_1)(D^2_x-D^2_y)\phi_{m.k_2}(\bm{r}_1)
\phi^{\dagger}_{m,k_2}(\bm{r}_2)(D^2_x-D^2_y)\phi_{l,k_1}(\bm{r}_2)\nonumber\\
&+e^{-i(\omega_m-\omega_l)(t_2-t_1)} \Theta(t_2-t_1) 
\phi_{m,k_2}^{\dagger}(\bm{r}_1)(D^2_x -D^2_y )\phi_{l.k_1}(\bm{r}_1)
\phi^{\dagger}_{l,k_1}(\bm{r}_2)(D^2_x-D^2_y )\phi_{m,k_2}(\bm{r}_2)\bigg],
\end{align}
in which $\omega_{m} = \bar \omega_c m+\alpha (m \bar \omega_c )^2, m \in {\mathbb Z}$ is the cyclotron energy of the composite fermion at the m-th Landau level. For $\nu=1/3$ filling, $p=1$. Thus the sum over $m$ simply becomes the sum over $m>0$ and we can set $l=0$ at the end of calculation. To proceed, we perform the Fourier transformation of the correlator.  Here we have denoted by
\begin{equation}
\phi_{l,k^{x}_{1}}(\bm{r}_1)=e^{ik^{x}_{1} x_1} \sqrt{\frac{1}{\sqrt{\pi}2^l l!}}e^{-(y_1+k^{x}_{1} )^2 /2} H_l (y_1+k^{x}_{1}), 
\end{equation}
the Landau wavefunctions in the $A_y=0$ gauge.
In Fourier space we find 
\begin{align}
&\langle  N_1 N_1 (\bm{q},\omega) \rangle
=C_{lm}\sum_{m} \int dk_i d^{2}x_i dy_i 
~\bigg[\frac{e^{i(x_2-x_1)(k^x_1-k^x_2+q_x)+iq_y(y_2-y_1)} }{\omega-(\omega_m-\omega_l)+i\epsilon} e^{(-1/2)((y_i+k^x_1)^2+(y_i+k^x_2)^2)}\nonumber\\
&H_l(y_1+k^x_1)(D^2_x-D^2_y )H_m(y_1+k^x_2)H_m(y_2+k^x_2)(D^2_x-D^2_y )H_l(y_2+k^x_1)\nonumber\\
&-\frac{e^{i(x_2-x_1)(-k^x_1+k^x_2+q_x)+iq_y(y_2-y_1)} }{\omega+(\omega_m-\omega_l)-i\epsilon} e^{(-1/2)((y_i+k^x_1)^2+(y_i+k^x_2)^2)}\nonumber\\
&H_m(y_1+k^x_2)(D^2_x-D^2_y )H_l(y_1+k^x_1)H_l(y_2+k^x_1)(D^2_x-D^2_y )H_m(y_2+k^x_2)\bigg],
\end{align}
\end{widetext}
where
\begin{equation}
C_{lm}=\frac{1}{2^{l+m} l! m! 2 \pi^2}
\end{equation}
We now change the variables to 
\begin{align}
\tilde{u_i}=&y_i+\frac{k^x_1+k^x_2}{2},\nonumber\\
\tilde{v}=&\frac{k^x_1-k^x_2}{2},\nonumber\\
u_1=&y_1+\frac{k^x_1+k^x_2}{2}+iq_y/2,\nonumber\\
u_2=&y_2+\frac{k^x_1+k^x_2}{2}-iq_y/2,\nonumber\\
v=&\tilde{v}-iq_y/2,\nonumber\\
v^*=&\tilde{v}+iq_y/2,
\end{align} 
and we integrate out $x_{i}$'s to obtain 
\begin{widetext}
\begin{align}
&\langle  N_1 N_1 (\bm{q},\omega)\rangle
=C_{lm}\sum_{m} \int du_i dv \nonumber\\
&[\frac{\delta(\tilde{v}+q_x/2)}{\omega-(\omega_m-\omega_l)+i\epsilon} e^{-u_i^2-2vv^*}
H_l(u_1+v)(D^2_x-D^2_y )H_m(u_1-v^*)
H_m(u_2-v)(D^2_x-D^2_y )H_l(u_2+v^*)\nonumber\\
&-\frac{\delta(\tilde{v}-q_x/2) }{\omega+(\omega_m-\omega_l)-i\epsilon} e^{-u_i^2-2vv^*}
H_m(u_1-v^*)(D^2_x-D^2_y )H_l(u_1+v)
H_l(u_2+v^*)(D^2_x-D^2_y )H_m(u_2-v)].
\end{align}
\end{widetext}
Since $-iD_x~H_m(u_1-v^*)=(u_1-v^*) H_m(u_1-v^*)$ (we choose the Landau gauge $D_x=\partial_x+i \bar b y, D_y=\partial_y$), we have the following. 
\begin{align}
-iD_x H_n=&1/2 H_{n+1}+nH_{N-1}\nonumber\\
-iD_y H_m(u_1-v^*)=& i(-1/2 H_{m+1}(u_1-v^*)\nonumber\\
+&mH_{m-1}(u_1-v^*))
\end{align}
In this way, the correlator can be simplified to the following expression
\begin{widetext}
\begin{align}
&\langle  N_1 N_1 (\bm{q},\omega) \rangle \nonumber\\
&=C_{lm}\sum_{m} \int du_i dv [\frac{\delta(\tilde{v}+q_x/2) }{\omega-(\omega_m-\omega_l)+i\epsilon} e^{-u_i^2-2vv^*}
m^2 H_{l+1} (u_1+v) H_{m-1}(u_1-v^*)H_{m-1}(u_2-v)H_{l-1}(u_2+v^*)\nonumber\\
&-\frac{\delta(\tilde{v}-q_x/2) }{\omega+(\omega_m-\omega_l)-i\epsilon} e^{-u_i^2-2vv^*}
m^2 H_{l+1}(u_1+v) H_{m-1}(u_1-v^*)H_{m-1}(u_2-v)H_{l-1}(u_2+v^*)]
\end{align}
\end{widetext}
We can use the expression for the inner product of the two Hermite polynomials which is written in terms of the Laguerre polynomials
\begin{align}
\int d u_1 e^{-u_i^2} H_{l}(u_1+v) H_{m}(u_1-v^*)=&\nonumber\\
2^m \sqrt{\pi} l! (v^*)^{m-l} &L_{l}^{m-l}(-2vv^*),
\end{align}
if $l$ is not larger than $m$. Here $v$ is related with $q_x, q_y$ after we integrate over $\tilde{v}$. $L_{l}^{m-l}(-2vv^*)$ is the polynomial of $q^2$ whose leading order is always a constant piece. We can always express the result by expanding it in terms of $\omega$ and $q$ by order. The leading order in $q$ and $\omega$ of $\langle  N_1 N_1 (q,\omega) \rangle $ (coming from $l=0,m=2$) includes a constant piece.
\begin{align}
\langle N_1 N_1 (q,\omega)\rangle=&\frac{1}{\pi(\omega-2\bar\omega_c)(1+4\alpha \bar \omega^2_c)}\nonumber\\
-&\frac{1}{\pi(\omega+2\bar\omega_c)(1+4\alpha \bar \omega^2_c)}+O(q^2)
\nonumber\\
=&\frac{4\bar\omega_c}{\pi(\omega^2-4\bar\omega_c^2)(1+4\alpha \bar \omega^2_c)}+O(q^2)
\end{align}
By dimensional analysis, we need to multiply $\bar \omega^2_c/ \bar l_b^2$ to the correlator to restore the coefficients by setting the magnetic length to be unity in the calculation. 
\begin{align}
\langle  N_1 N_1 (q,\omega) \rangle =&\frac{4\bar \omega_c^3}{\bar l_b^2 \pi(\omega^2-4\bar \omega_c^2)(1+4\alpha \bar \omega^2_c)}+O(q^2)
\nonumber\\
=&-\frac{\bar \omega_c}{\bar l_b^2 \pi(1+4\alpha \bar \omega^2_c)}+O(q^2)+O(\omega^2)
\end{align}
The first term will contribute to the mass term of the nematic order parameters.

The anti-symmetric part, proportional to  $ M_{1} \partial_{0}M_{2}$, of the correlator can be calculated in the same way. The full result for the correlator is
\begin{widetext}
\begin{align}
&\langle N_1 N_2 (\bm{q},\omega)\rangle
=C_{lm}\sum_{m} \int du_i dv \nonumber\\
&[\frac{\delta(\tilde{v}+q_x/2)}{\omega-(\omega_m-\omega_l)+i\epsilon} e^{-u_i^2-2vv^*}
H_l(u_1+v)(D^2_x-D^2_y )H_m(u_1-v^*)
H_m(u_2-v)(D^2_x-D^2_y )H_l(u_2+v^*)\nonumber\\
&-\frac{\delta(\tilde{v}-q_x/2) }{\omega+(\omega_m-\omega_l)-i\epsilon} e^{-u_i^2-2vv^*}
H_m(u_1-v^*)(D^2_x-D^2_y )H_l(u_1+v)
H_l(u_2+v^*)(D^2_x-D^2_y )H_m(u_2-v)]\nonumber\\
&=C_{lm}\sum_{m} \int du_i dv [\frac{\delta(\tilde{v}+q_x/2) }{\omega-(\omega_m-\omega_l)+i\epsilon} e^{-u_i^2-2vv^*}
\frac{m^2}{i} H_{l+1} (u_1+v) H_{m-1}(u_1-v^*)H_{m-1}(u_2-v)H_{l-1}(u_2+v^*)\nonumber\\
&+\frac{\delta(\tilde{v}-q_x/2) }{\omega+(\omega_m-\omega_l)-i\epsilon} e^{-u_i^2-2vv^*}
\frac{m^2}{i} H_{l+1}(u_1+v) H_{m-1}(u_1-v^*)H_{m-1}(u_2-v)H_{l-1}(u_2+v^*)]
\end{align}
\end{widetext}
The leading order behavior  in $q$ and $\omega$ comes from the term with $l=0$ and $m=2$. Within this approximation, the leading low frequency and low momenta behavior of the correlator is
\begin{equation}
\langle N_1 N_2 (q,\omega)\rangle=\frac{2\omega}{i\pi(\omega^2-4\bar \omega_c^2)(1+4\alpha \bar \omega^2_c)^2}+O(q^2)
\end{equation}
Again, we now multiply the factor $\bar \omega_c^2/\bar l^2_b$ to restore the units properly to find the result
\begin{equation}
\langle N_1 N_2 (q,\omega)\rangle=i\frac{\omega }{2 \bar l^2_b \pi(1+4\alpha \bar \omega^2_c)^2}+O(q^2)+O(\omega^2) 
\end{equation}
The coefficient of the leading term   is the Hall viscosity  of the integer quantum Hall state. 

\section{Calculation of the Mixed Correlators of Nematic and Gauge Fields}
\label{app:mixed-correlators}

\begin{figure}[hbt]
   \centering
  \subfigure[]{\includegraphics[width=0.3\textwidth]{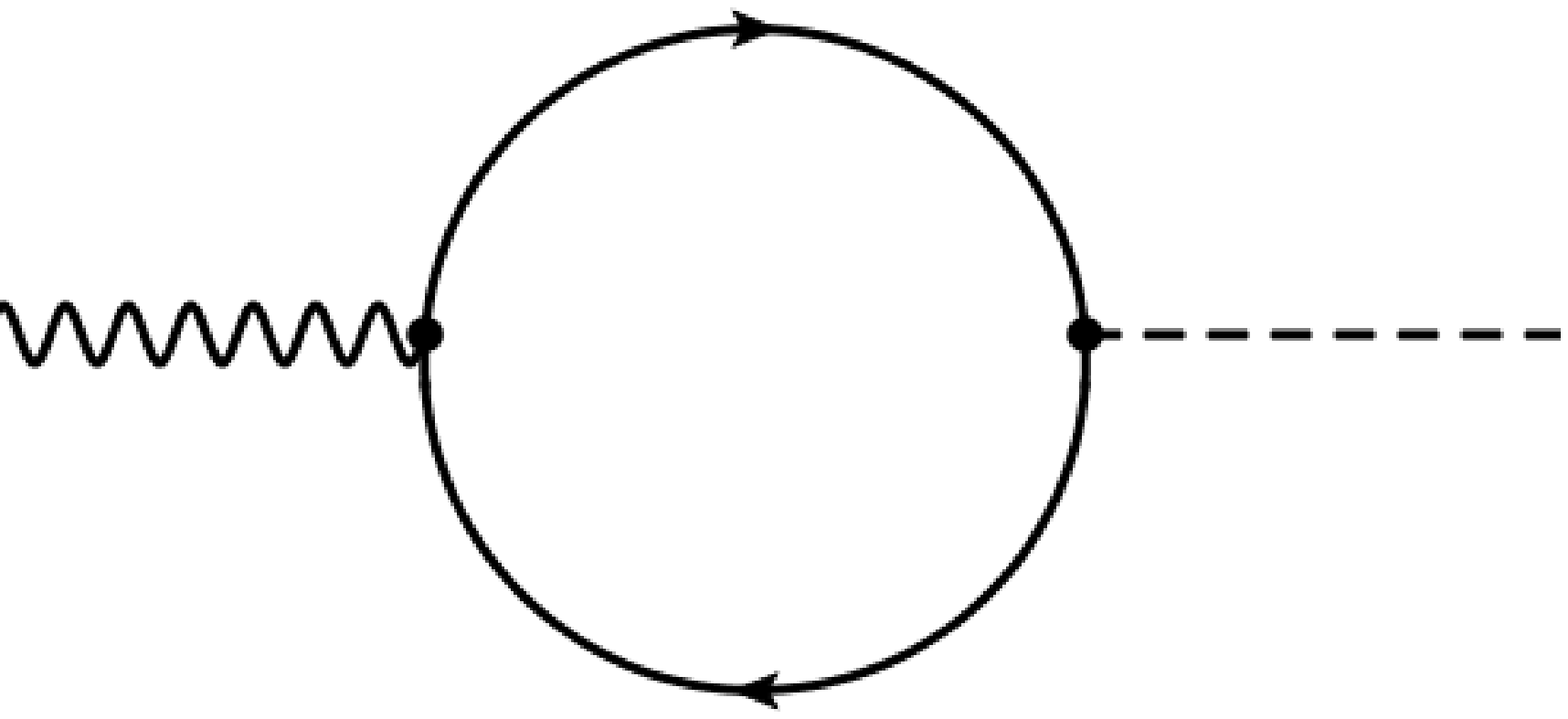}}
    \subfigure[]{\includegraphics[width=0.3\textwidth]{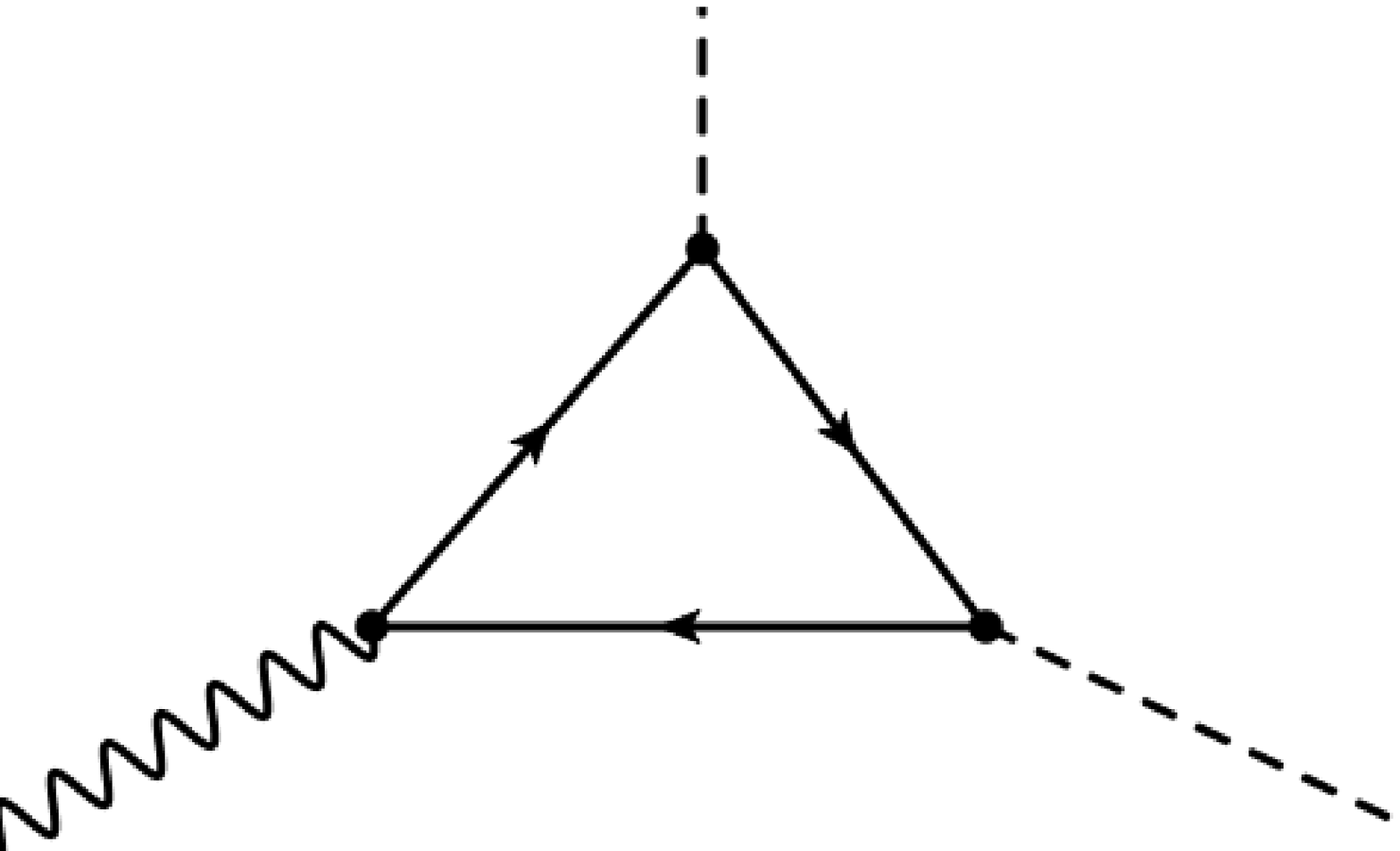}}
\subfigure[]{\includegraphics[width=0.35\textwidth]{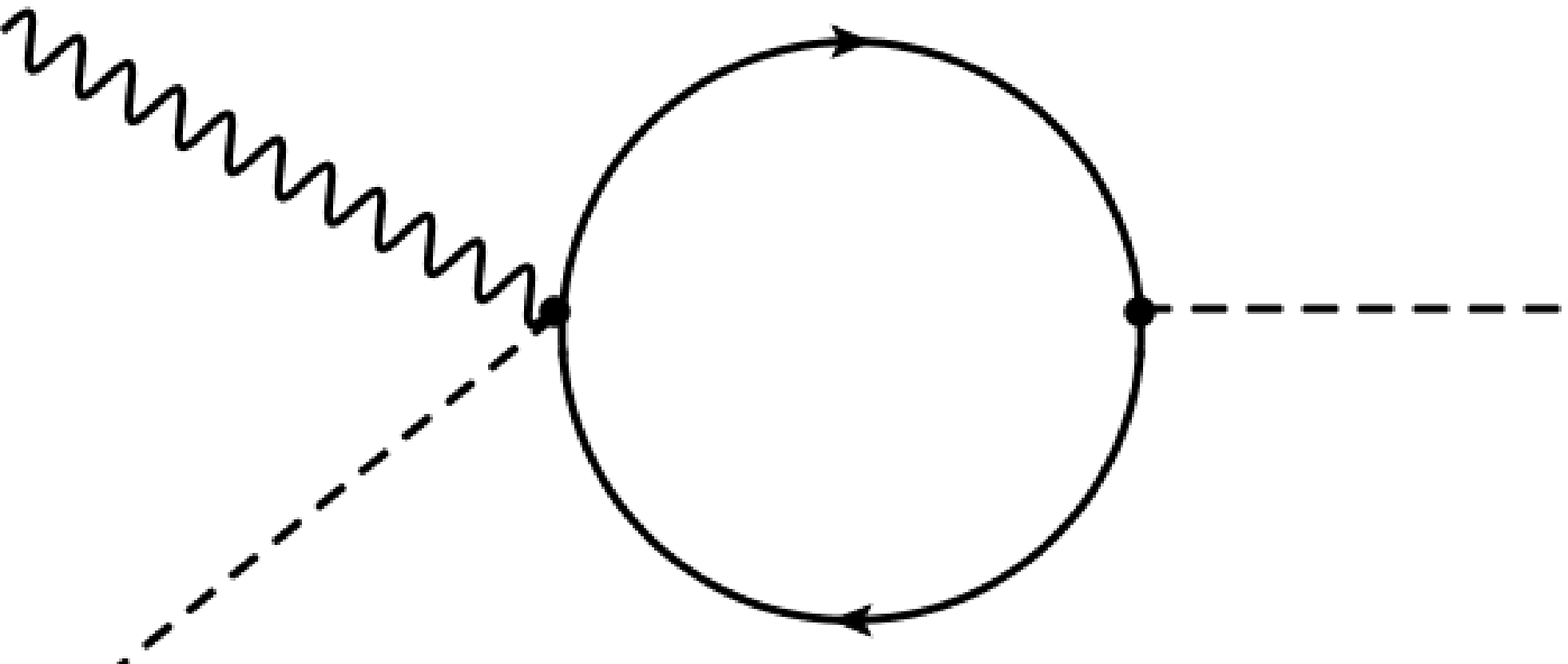}}
\caption{Diagrams contributing to the Wen-Zee term. Here the wiggly line represents the gauge field $a_{\mu}$ and the dotted  line represents the nematic field $M_{i}$. The  thick line represents the composite fermion propagator.}
\label{fig:wenzee}
\end{figure}

The nematic-gauge coupling term could be obtained in a similar way
\begin{widetext}
\begin{align}
& \langle N_1 (\bm{r_1}, t_1) j_0 (\bm{r_2}, t_2) \rangle=-i\langle \mathcal{T} \Psi^{\dagger}(\bm{r}_1,t_1)(D^2_x-D^2_y )\Psi(\bm{r}_1,t_1)
\Psi^{\dagger}(\bm{r}_2,t_2)\Psi(\bm{r}_2,t_2)\rangle \nonumber\\
&=C_{lm}\sum_{m} \int du_i dv \nonumber\\
&[\frac{\delta(\tilde{v}+q_x/2)}{\omega-(\omega_m-\omega_l)+i\epsilon} e^{-u_i^2-2vv^*}
H_l(u_1+v)(D^2_x-D^2_y )H_m(u_1-v^*)
H_m(u_2-v)H_l(u_2+v^*)\nonumber\\
&-\frac{\delta(\tilde{v}-q_x/2) }{\omega+(\omega_m-\omega_l)-i\epsilon} e^{-u_i^2-2vv^*}
H_m(u_1-v^*)(D^2_x-D^2_y )H_l(u_1+v)
H_l(u_2+v^*)H_m(u_2-v)]\nonumber\\
&=C_{lm}\sum_{m} \int du_i dv [\frac{\delta(\tilde{v}+q_x/2) }{\omega-(\omega_m-\omega_l)+i\epsilon} e^{-u_i^2-2vv^*}
m H_{l+1} (u_1+v) H_{m-1}(u_1-v^*)H_{m}(u_2-v)H_{l}(u_2+v^*)\nonumber\\
&-\frac{\delta(\tilde{v}-q_x/2) }{\omega+(\omega_m-\omega_l)-i\epsilon} e^{-u_i^2-2vv^*}
m H_{l+1}(u_1+v) H_{m-1}(u_1-v^*)H_{m}(u_2-v)H_{l}(u_2+v^*)]
\end{align}
\end{widetext}
The leading order term is $\propto p^2$, which comes from the contribution of $(l=0,m=2),(l=0,m=1)$, and yields the expression,
\begin{equation}
 \langle N_1(p)  j_0(-p)  \rangle=\frac{1}{2\pi}(\frac{2}{1+\alpha \bar \omega^2_c} -\frac{2}{2+8\alpha \bar\omega^2_c})(p_x^2-p_y^2)
\end{equation}
Following the above calculation, we can obtain other linear coupling terms between the nematic field and gauge field,
\begin{align}
 \langle N_2(p)  j_0(-p)  \rangle=&\frac{1}{2\pi}(\frac{2}{1+\alpha \bar \omega^2_c} -\frac{2}{2+8\alpha \bar\omega^2_c}) 2 p_x p_y \nonumber\\
 \langle N_1(p)  j_x(-p)  \rangle=&\frac{1}{2\pi}(\frac{2}{1+\alpha \bar \omega^2_c} -\frac{2}{2+8\alpha \bar\omega^2_c}) \omega p_x \nonumber\\
 \langle N_1(p)  j_y(-p)  \rangle=&\frac{1}{2\pi}(\frac{2}{1+\alpha \bar \omega^2_c} -\frac{2}{2+8\alpha \bar\omega^2_c})(-\omega p_y)\nonumber\\
 \langle N_2(p)  j_x(-p)  \rangle=&\frac{1}{2\pi}(\frac{2}{1+\alpha \bar \omega^2_c} -\frac{2}{2+8\alpha \bar\omega^2_c}) \omega p_y\nonumber\\
 \langle N_2(p)  j_y(-p)  \rangle=&\frac{1}{2\pi}(\frac{2}{1+\alpha \bar \omega^2_c} -\frac{2}{2+8\alpha \bar\omega^2_c}) \omega p_x
\end{align}

These terms contribute partly to the Wen-Zee coupling. For a complete expression of the Wen-Zee term, we also needs to evaluate the correlator between two nematic field and one gauge field
\begin{widetext}
\begin{align}
& \langle N_1 (\bm{r_1}, t_1) N_2 (\bm{r_2}, t_2) j_0 (\bm{r_3}, t_3) \rangle=\nonumber\\
&-\langle \mathcal{T} \Psi^{\dagger}(\bm{r}_1,t_1)(D^2_x-D^2_y )\Psi(\bm{r}_1,t_1)\Psi^{\dagger}(\bm{r}_2,t_2)(D_xD_y+D_y D_x)\Psi(\bm{r}_2,t_2)
\Psi^{\dagger}(\bm{r}_3,t_3)\Psi(\bm{r}_3,t_3)\rangle\\
& \langle N_1  N_2 j_0\rangle(q,p) = \langle N_1 (\bm{r_1}, t_1) N_2 (\bm{r_2}, t_2) j_0 (\bm{r_3}, t_3) \rangle \exp(-i\omega(t_2-t_1)-i\omega_0(t_3-t_2))\exp(iq (r_2-r_1)+ip(r_3-r_2))
\end{align}
By redefining the variables,
\begin{align}
&u_1=y_1+\frac{k_1^x+k_2^x}{2}+iq^y/2, ~v=\frac{k_1^x-k_2^x}{2}-iq^y/2,~\tilde{v}=\frac{k_1^x-k_2^x}{2};\nonumber\\
&u_2=y_2+\frac{k_2^x+k_3^x}{2}-i(q^y-p^y)/2, ~v_0=\frac{k_2^x-k_3^x}{2}+i(q^y-p^y)/2,~\tilde{v_0}=\frac{k_2^x-k_3^x}{2};\nonumber\\
&u_3=y_3+\frac{k_1^x+k_3^x}{2}-ip^y/2, ~v+v_0=\frac{k_1^x-k_3^x}{2}+ip^y/2, ~\tilde{v}+~\tilde{v_0}=\frac{k_1^x-k_3^x}{2};
\end{align}
we can write the time ordered correlator (for $t_1>t_2>t_3$) as
\begin{align}
 &\langle N_1  N_2 j_0\rangle(q,p) =exp(-u_i^2-vv^*-v_0 v^*_0-(v_0+v)(v^*+ v^*_0))~exp(-iq/2\wedge p /2)\nonumber\\
&C_{lmn}\sum_{m,n} \int du_i dv dv_0 ~\frac{\delta(\tilde{v}+q^x/2)\delta(\tilde{v_0}+p^x/2)}{[\omega-(\omega_m-\omega_l)+i\epsilon][\omega_0-(\omega_n-\omega_l)+i\epsilon]} \nonumber\\
&H_l(u_1+v)(D^2_x-D^2_y )H_m(u_1-v^*)H_m(u_2+v_0)(D_xD_y+D_y D_x)H_n(u_2-v_0^*)
H_n(u_3-v^*-v_0^*)H_l(u_3+v+v_0)\nonumber\\
&=exp(-u_i^2-vv^*-v_0 v^*_0-(v_0+v)(v^*+ v^*_0))~exp(-iq/2\wedge p/2)\nonumber\\
&C_{lmn}\sum_{m,n} \int du_i dv dv_0 ~\frac{\delta(\tilde{v}+q^x/2)\delta(\tilde{v_0}+p^x/2)}{[\omega-(\omega_m-\omega_l)+i\epsilon][\omega_0-(\omega_n-\omega_l)+i\epsilon]} \nonumber\\
&[H_{l+1}(u_1+v)m H_{m-1}(u_1-v^*)H_{m+1}(u_2+v_0)(in)H_{n-1}(u_2-v_0^*)
H_n(u_3-v^*-v_0^*)H_l(u_3+v+v_0)\nonumber\\
&+H_{l+1}(u_1+v)m H_{m-1}(u_1-v^*)H_{m-1}(u_2+v_0)(-im)H_{n+1}(u_2-v_0^*)
H_n(u_3-v^*-v_0^*)H_l(u_3+v+v_0)]
\end{align}
\end{widetext}
For this three-point correlator, there always exists an antisymmetric phase factor $\exp(-iq/2 \wedge p/2)$(known as the Moyal phase, see e.g. Ref. \cite{fradkin-2002}) which is responsible for the Wen-Zee response. 

The leading order contribution for three point time ordered correlator (at $t_1>t_2>t_3$) comes from the
choice of $[l=0,n=0,m=2]$, thus
\begin{align}
 \langle N_1  N_2 j_0\rangle(q,p)|_{t_1>t_2>t_3}=&\nonumber\\
 &\frac{-q_xp_y+q_yp_x}{4\pi(\omega-(\omega_2-\omega_0))(\omega_0)}+....
\end{align}
In a similar way, we also have
\begin{widetext}
\begin{align}
 &\langle N_2  N_1 j_0\rangle(q,p)|_{t_1>t_2>t_3} =\exp(-u_i^2-vv^*-v_0 v^*_0-(v_0+v)(v^*+ v^*_0)) \exp(-iq/2\wedge p/2)\nonumber\\
&C_{lmn}\sum_{m,n} \int du_i dv dv_0 ~\frac{\delta(\tilde{v}+q^x/2)\delta(\tilde{v_0}+p^x/2)}{[\omega-(\omega_m-\omega_l)+i\epsilon][\omega_0-(\omega_n-\omega_l)+i\epsilon]} \nonumber\\
&H_l(u_1+v)(D_xD_y+D_y D_x)H_m(u_1-v^*)H_m(u_2+v_0)(D^2_x-D^2_y )H_n(u_2-v_0^*)
H_n(u_3-v^*-v_0^*)H_l(u_3+v+v_0)\nonumber\\
&=\exp(-u_i^2-vv^*-v_0 v^*_0-(v_0+v)(v^*+ v^*_0)) \exp(-iq/2\wedge p/2)\nonumber\\
&C_{lmn}\sum_{m,n} \int du_i dv dv_0 ~\frac{\delta(\tilde{v}+q^x/2)\delta(\tilde{v_0}+p^x/2)}{[\omega-(\omega_m-\omega_l)+i\epsilon][\omega_0-(\omega_n-\omega_l)+i\epsilon]} \nonumber\\
&[H_{l+1}(u_1+v)(im) H_{m-1}(u_1-v^*)H_{m+1}(u_2+v_0)(n)H_{n-1}(u_2-v_0^*)
H_n(u_3-v^*-v_0^*)H_l(u_3+v+v_0)\nonumber\\
&+H_{l+1}(u_1+v)(im) H_{m-1}(u_1-v^*)H_{m-1}(u_2+v_0)(m)H_{n+1}(u_2-v_0^*)
H_n(u_3-v^*-v_0^*)H_l(u_3+v+v_0)]
\end{align}
\end{widetext}
In the leading order,
\begin{align}
 &\langle N_2  N_1 j_0\rangle(q,q_0)|_{t_1>t_2>t_3}=\frac{q_xp_y-q_yp_x}{4\pi(\omega-(\omega_2-\omega_0))(\omega_0)}+....
\end{align}
The other time ordered correlator can be obtained in the similar way which finally gives Wen-Zee coupling.
Finally, we have,
\begin{align}
\langle N_i  N_j  j_{\mu}\rangle (q,p)=\frac{\epsilon^{ij} \epsilon^{\lambda \nu \mu}  p_{\lambda} q_{\nu}}{(1+4\alpha \bar \omega_c^2)^2 4\pi}
\end{align}

\section{Proof of gauge invariance at the RPA level}
\label{app:gauge-invariance}

To calculate the collective excitation of the Nematic FQH state, we first treat the Quadrupolar interaction perturbatively in the RPA level and then integrate out the gauge fluctuation. During the RPA procedure, we only keep the reducible diagrams for the infinite geometric series.  Here, we would proof that the polarization tensor of  such RPA level correction is gauge invariant.
The polarization at RPA level has the form
\begin{align}
\Pi^{RPA}_{ij}=&\Pi^{0}_{ij}+(2F_2 m_e^2)\sum_{a,b}\langle  j_i N_a \rangle \langle  N_b j_j \rangle \nonumber\\
& +(2F_2m_e^2)^2\sum_{a,b}\langle  j_i N_a \rangle \langle  N_a N_b\rangle \langle  N_b j_j \rangle +\cdots \nonumber\\
=&\Pi^{0}_{ij}+\frac{(2F_2m_e^2)\langle  j_i N_a \rangle \langle  N_b j_j \rangle}{1-(2F_2m_e^2) \langle  N_a N_b \rangle}
\end{align}
To proof its gauge invariance, we only need to proof
\begin{align}
p_{\mu}\Pi^{RPA}_{\mu \nu}(\omega,p)=0
\end{align}
Or if write it in the real space,
\begin{align}
\partial_{\mu} \Pi^{RPA}_{\mu \nu} (x,y)=0
\end{align}
It is obvious that $p_{\mu}\Pi^{0}_{\mu\nu}=0$. Thus we only need to prove the gauge invariance of $p_{\mu} \langle  j_{\mu} N_a \rangle=0$.
\begin{widetext}
\begin{align}
&2 \partial_{r^1_i}\langle  j_{i} (r_1) N_1(r_2) \rangle =i\partial_{r^1_i}\langle D_i^{\dagger}\Psi^{\dagger}(r_1)\Psi(r_1)\Psi^{\dagger}(r_2)(D_x^2-D_y^2)\Psi(r_2)\rangle 
-i\partial_{r^1_i}\langle\Psi^{\dagger}(r_1)D_i\Psi(r_1)\Psi^{\dagger}(r_2)(D^2_x-D^2_y )\Psi(r_2)\rangle \nonumber\\
&=i\langle\partial_{r^1_i} D_i^{\dagger}\Psi^{\dagger}(r_1)\Psi(r_1)\Psi^{\dagger}(r_2)(D^2_x-D^2_y )\Psi(r_2)\rangle +i\langle D_i^{\dagger}\Psi^{\dagger}(r_1)\partial_{r^1_i}\Psi(r_1)\Psi^{\dagger}(r_2)(D_x^2-D_y^2)\Psi(r_2)\rangle \nonumber\\
&-i\langle\partial_{r^1_i}\Psi^{\dagger}(r_1)D_i\Psi(r_1)\Psi^{\dagger}(r_2)(D_x^2-D_y^2)\Psi(r_2)\rangle 
-i\langle\Psi^{\dagger}(r_1)\partial_{r^1_i}D_i\Psi(r_1)\Psi^{\dagger}(r_2)(D_x^2-D_y^2)\Psi(r_2)\rangle \nonumber\\
&=i\langle D_i^{\dagger} D_i^{\dagger}\Psi^{\dagger}(r_1)\Psi(r_1)\Psi^{\dagger}(r_2)(D_x^2-D_y^2)\Psi(r_2)\rangle 
+i\langle  D_i^{\dagger}\Psi^{\dagger}(r_1)D_i \Psi(r_1)\Psi^{\dagger}(r_2)(D_x^2-D_y^2)\Psi(r_2)\rangle \nonumber\\
&-i\langle D_i^{\dagger}\Psi^{\dagger}(r_1)D_i\Psi(r_1)\Psi^{\dagger}(r_2)(D_x^2-D_y^2)\Psi(r_2)\rangle -i\langle \Psi^{\dagger}(r_1)D_i D_i\Psi(r_1)\Psi^{\dagger}(r_2)(D_x^2-D_y^2)\Psi(r_2)\rangle \nonumber\\
&=i\langle D_i^{\dagger} D_i^{\dagger}\Psi^{\dagger}(r_1)\Psi(r_1)\Psi^{\dagger}(r_2)(D_x^2-D_y^2)\Psi(r_2)\rangle -i\langle \Psi^{\dagger}(r_1)D_i D_i\Psi(r_1)\Psi^{\dagger}(r_2)(D_x^2-D_y^2)\Psi(r_2)\rangle \nonumber\\
&\partial_{r^1_0}\langle  j_{0} (r_1) N_1(r_2) \rangle=-\langle \partial_{r^1_0} \Psi^{\dagger}(r_1)\Psi(r_1)\Psi^{\dagger}(r_2)(D_x^2-D_y^2)\Psi(r_2)\rangle -\langle \Psi^{\dagger}(r_1)\partial_{r^1_0} \Psi(r_1)\Psi^{\dagger}(r_2)(D_x^2-D_y^2)\Psi(r_2)\rangle 
\end{align}
In all we have,
\begin{align}
& -i \partial_{r^1_{\mu}}\langle  j_{\mu} (r_1) N_1(r_2) \rangle =\langle \Psi(r_1)\Psi^{\dagger}(r_2)(D_x^2-D_y^2)_{r_2} \Psi(r_2)\Psi^{\dagger}(r_1)(-i\partial_0+
\frac{{{\bm D}^\dagger}^2}{2}+\mu)_{r_1}\rangle \nonumber\\
&-\langle (i\partial_0+\frac{{\bm D}^2}{2}+\mu)_{r_1}\Psi(r_1)\Psi^{\dagger}(r_2)(D_x^2-D_y^2)_{r_2} \Psi(r_2)\Psi^{\dagger}(r_1)\rangle \nonumber\\
&=\langle G(r_1,r_2)(D_x^2-D_y^2)_{r_2} G(r_2,r_1)(-i\partial_0+\frac{{{\bm D}^\dagger}^2}{2}+\mu)_{r_1}\rangle 
-\langle (i\partial_0+\frac{{\bm D}^2}{2}+\mu)_{r_1}G(r_1,r_2)(D_x^2-D_y^2)_{r_2} G(r_2,r_1)\rangle 
\end{align}
\end{widetext}
Recall that 
the Green function has the property,
\begin{align}
\Big(i\partial_0+\frac{{\bm D}^2}{2}+\mu \Big)_{r_1}&G(r_1,r_2)=\nonumber\\
&\Big(i\partial_0+\frac{{\bm D}^2}{2}+\mu \Big)_{r_1}\langle\Psi(r_1)\Psi^{\dagger}(r_2)\rangle
\nonumber\\
&=\delta(r_1-r_2)
\end{align}
and similarly for the adjoint. 
Thus we have,
\begin{equation}
 -i \partial_{r^1_{\mu}}\langle  j_{\mu} (r_1) N_1(r_2) \rangle =0
\end{equation}
Thus, we had shown that the polarization tensor at the RPA level is gauge invariant.

\section{Nematic collective excitations}
\label{app:collexcitation}

To obtain the collective excitations of the nematic FQH state, we need to calculate the polarization tensor$ K_{ij}$ for external electromagnetic gauge field. The poles in $K_{ij}$ gives the spectrum of the excitations. 
\begin{align}
&K_{00}=q^2 K_0, \nonumber\\
&K_{0i}=\omega q_i K_0+i \epsilon_{ik} q_k K_1, \nonumber\\
&K_{i0}=\omega q_i K_0-i \epsilon_{ik} q_k K_1, \nonumber\\
&K_{ij}=\omega^2 \delta_{ij} K_0-i\epsilon_{ij} \omega K_1+(q^2 \delta_{ij}-q_iq_j)K_2, \nonumber\\
&K_0=-\frac{\Pi_0}{16\pi^2 D},\nonumber\\
&K_1=\frac{1}{4\pi }+\frac{\Pi_1+\frac{1}{4\pi}}{16\pi^2 D}+\frac{V(q)\Pi_0 q^2}{64\pi^3 D},\nonumber\\
&K_2=\frac{\Pi_2}{16\pi^2 D}+\frac{V(q)(\omega^2\Pi_0^2-\Pi_1^2)}{ D}+\frac{V(q)\Pi_0 \Pi_2 q^2}{ D},\nonumber\\
&D=\Pi_0^2 \omega^2-(\Pi_1+\theta)^2+\Pi_0(\Pi_2-\frac{V(q)}{16\pi^2})q^2,\nonumber\\
&\theta=\frac{1}{4\pi}.
\end{align}
We solve $D(q,\omega)=0$ to find the poles of $K_{\mu\nu}$.
\begin{align}
\label{K}
\Pi_0^2 \omega^2-(\Pi_1+\theta)^2+\Pi_0(\Pi_2-\frac{V(q)}{16\pi^2})q^2=0
\end{align}
As the left part of the equation \ref{K} involves a sum of infinite numbers of polynomials, it is impossible to solve it exactly. What we can try to do instead is to assume dispersion $\omega=\omega_1+\alpha_n q^{2n}$ and find the solution asymptotically near zero momentum. We only keep the lowest terms in $q$ and $\omega$. As we can see, the first two terms have leading order $O(1)$, while the last term have the leading order $O(q^2)$. To solve this equation, we have to subtract the constant piece from the first two terms and set them as zero. 
\begin{align}
&|\frac{\bar \omega_c' \omega_1}{\omega_1^2-\bar \omega_c'^2}+\frac{c_1 t^2 \omega_1}{\alpha}|=|\frac{\bar \omega_c'^2}{\omega_1^2-\bar \omega_c'^2}+\frac{t^2 c_2}{\alpha}+2\pi \theta |
\end{align}
This gives us the solution for the lowest excitation in the nematic FQH state, which is the GMP mode.

%


\end{document}